%% file: main.tex
\documentclass[fleqn,usenatbib]{mnras}

\usepackage{newtxtext,newtxmath}
\usepackage[T1]{fontenc}

\DeclareRobustCommand{\VAN}[3]{#2}
\let\VANthebibliography\thebibliography
\def\thebibliography{\DeclareRobustCommand{\VAN}[3]{##3}\VANthebibliography}

\usepackage{graphicx}	% Including figure files
\usepackage{amsmath}	% Advanced maths commands
\usepackage{amssymb}	% Extra maths symbols

\usepackage{xcolor}

\graphicspath{{./}{./figures/}}

\newcommand*{\mysub}[2]{\ensuremath{#1_{\mathrm{#2}}}}
\newcommand*{\unit}[1]{\ensuremath{\mathrm{\, #1}}}
\newcommand*{\phmin}{\hspace{1.9ex}} % phantom minus sign

\newcommand*{\E}[1]{\ensuremath{\times 10^{#1}}}
\newcommand*{\ltsim}{\ {\raise-.75ex\hbox{$\buildrel<\over\sim$}}\ }
\newcommand*{\gtsim}{\ {\raise-.75ex\hbox{$\buildrel>\over\sim$}}\ }
\newcommand*{\dnorm}{\ensuremath{\mathcal{N}}}
\newcommand*{\dunif}{\ensuremath{\mathcal{U}}}

\newcommand*{\Msun}{\ensuremath{M_{\sun}}}

\newcommand*{\cm}{\unit{cm}}

\newcommand*{\Omegam}{\mysub{\Omega}{m}}
\newcommand*{\Omegab}{\mysub{\Omega}{b}}
\newcommand*{\Omegak}{\mysub{\Omega}{k}}
\newcommand*{\Omegal}{\mysub{\Omega}{\Lambda}}
\newcommand*{\Omegade}{\mysub{\Omega}{DE}}
\newcommand*{\LCDM}{\ensuremath{\Lambda}CDM}

\newcommand*{\fgas}{\mysub{f}{gas}}

\newcommand*{\Ugas}{\mysub{\Upsilon}{gas}}
\newcommand*{\Mgas}{\mysub{M}{gas}}
\newcommand*{\Mtot}{\mysub{M}{tot}}
\newcommand*{\rhocr}{\mysub{\rho}{cr}}

\newcommand*{\NH}{\mysub{N}{H}}

\newcommand*{\Chandra}{{\it Chandra}}

\newcommand*{\Planck}{{\it Planck}}

\defcitealias{Mantz1402.6212}{M14}
\newcommand*{\mamrakls}{\citetalias{Mantz1402.6212}}
\defcitealias{Mantz1502.06020}{M15}
\newcommand*{\mamslu}{\citetalias{Mantz1502.06020}}
\defcitealias{Applegate1509.02162}{A16}
\newcommand*{\amalmhkbers}{\citetalias{Applegate1509.02162}}
\defcitealias{Navarro9611107}{NFW}
\newcommand*{\NFW}{\citetalias{Navarro9611107}}

\title[Cosmological Constraints from \fgas{}]{%
  Cosmological Constraints from Gas Mass Fractions of Massive, Relaxed Galaxy Clusters
}

\author[A.\ B.\ Mantz et al.]{Adam B.\ Mantz,$^{1}$\thanks{Corresponding author e-mail: \href{mailto:amantz@stanford.edu}{\tt amantz@stanford.edu}}
  R. Glenn Morris,$^{1,2}$
  Steven W. Allen,$^{1,2,3}$
  Rebecca E. A. Canning,$^{1,4}$
  Lucie Baumont,$^{5}$\newauthor
  Bradford Benson,$^{6,7,8}$
  Lindsey E. Bleem,$^{8,9}$
  Steven R. Ehlert,$^{10}$
  Benjamin Floyd,$^{11}$
  Ricardo Herbonnet,$^{5}$\newauthor
  Patrick L. Kelly,$^{12}$
  Shuang Liang,$^{5}$
  Anja von der Linden,$^{5}$
  Michael McDonald,$^{13}$
  David A. Rapetti,$^{14,15,16}$\newauthor
  Robert W. Schmidt,$^{17}$
  Norbert Werner,$^{18}$
  and
  Adam Wright$^{19}$
  \smallskip
  \\$^{1}$Kavli Institute for Particle Astrophysics and Cosmology, Stanford University, 452 Lomita Mall, Stanford, CA 94305, USA
  \\$^{2}$SLAC National Accelerator Laboratory, 2575 Sand Hill Road, Menlo Park, CA  94025, USA
  \\$^{3}$Department of Physics, Stanford University, 382 Via Pueblo Mall, Stanford, CA 94305, USA 
  \\$^{4}$Institute of Cosmology and Gravitation, University of Portsmouth, Portsmouth, PO1 3FX, UK 
  \\$^{5}$Department of Physics and Astronomy, Stony Brook University, Stony Brook, NY 11794, USA 
  \\$^{6}$Fermi National Accelerator Laboratory, Batavia, IL
  60510-0500, USA
  \\$^{7}$Department of Astronomy and Astrophysics, University
  of Chicago, 5640 S Ellis Ave, Chicago, IL 60637, USA
  \\$^{8}$Kavli Institute for Cosmological Physics, University of
  Chicago, 5640 S Ellis Ave, Chicago, IL 60637, USA
  \\$^{9}$HEP Division, Argonne National Laboratory, Argonne, IL 60439, USA 
  \\$^{10}$Marshall Space Flight Center, Huntsville, AL 35812, USA
  \\$^{11}$Department of Physics and Astronomy, University of Missouri--Kansas City, 5110 Rockhill Road, Kansas City, MO 64110, USA 
  \\$^{12}$School of Physics and Astronomy, University of Minnesota, 116 Church St SE, Minneapolis, MN 55455
  \\$^{13}$Kavli Institute for Astrophysics and Space Research, Massachusetts Institute of Technology, 77 Massachusetts Avenue, \\Cambridge, MA 02139, USA
  \\$^{14}$NASA Ames Research Center, Moffett Field, CA 94035, USA
  \\$^{15}$Research Institute for Advanced Computer Science, Universities Space Research Association, Mountain View, CA 94043, USA
  \\$^{16}$Center for Astrophysics and Space Astronomy, Department of Astrophysical and Planetary Sciences, University of Colorado, Boulder, CO 80309, USA
  \\$^{17}$Astronomisches Rechen-Institut, Zentrum f\"ur Astronomie der Universit\"at Heidelberg, M\"onchhofstrasse 12-14, D-69120 Heidelberg, Germany
  \\$^{18}$Department of Theoretical Physics and Astrophysics, Faculty of Science, Masaryk University, Kotl\'a\v{r}sk\'a 2, Brno, 611 37, Czech Republic 
  \\$^{19}$Department of Physics and Chemistry, Milwaukee School of Engineering, 432 E Kilbourn Ave, Milwaukee, WI 53202, USA
}

\date{Accepted 2021 November 16. Received 2021 November 5; in original form 2021 August 6}

\pubyear{2021}

\begin{document}
\label{firstpage}
\pagerange{\pageref{firstpage}--\pageref{lastpage}}
\maketitle

\begin{abstract}
  We present updated cosmological constraints from measurements of the gas mass fractions (\fgas) of massive, dynamically relaxed galaxy clusters. 
  Our new data set has greater leverage on models of dark energy, thanks to the addition of the Perseus Cluster at low redshifts, two new clusters at redshifts $z\gtsim1$, and significantly longer observations of four clusters at $0.6<z<0.9$.
  Our low-redshift ($z<0.16$) \fgas{} data, combined with the cosmic baryon fraction measured from the cosmic microwave background (CMB), imply a Hubble constant of $h=0.722\pm0.067$.
  Combining the full \fgas{} data set with priors on the cosmic baryon density and the Hubble constant, we constrain the dark energy density to be $\Omegal=0.865\pm0.119$ in non-flat \LCDM{} (cosmological constant) models, and its equation of state to be $w=-1.13_{-0.20}^{+0.17}$ in flat, constant-$w$ models, respectively 41 and 29 per cent tighter than our previous work, and comparable to the best constraints available from other probes.
  Combining \fgas, CMB, supernova, and baryon acoustic oscillation data, we also constrain models with global curvature and evolving dark energy.
  For the massive, relaxed clusters employed here, we find the scaling of \fgas{} with mass to be consistent with a constant, with an intrinsic scatter that corresponds to just $\sim3$ per cent in distance.
\end{abstract}


%%%%%%%%%%%%%%%%%%%%%%%%%%%%%%%%%%%%%%%%%%%%%%%%%%

%%%%%%%%%%%%%%%%% BODY OF PAPER %%%%%%%%%%%%%%%%%%

\section{Introduction}

For massive clusters of galaxies, whose internal dynamics are dominated by gravity, the mass of the intracluster medium (ICM) correlates tightly with total mass (\citealt{Borgani0906.4370}, and references therein).
Furthermore, the relationship between the gas-mass fraction of these clusters, $\fgas=\Mgas/\Mtot$, and the cosmic baryon mass fraction, $\Omegab/\Omegam$, is both straightforward to predict from simulations of cosmic structure formation and minimally sensitive to cosmological modeling assumptions (\citealt{Eke9708070, Kay0407058, Crain0610602}; \citealt*{Nagai0609247}; \citealt{Young1007.0887, Battaglia1209.4082, Planelles1209.5058, Le-Brun1312.5462, Le-Brun1606.04545, Barnes1607.04569, Henden1911.12367, Singh1911.05751}).
Precise measurements of $\fgas$ from X-ray data thus provide a route to constraining cosmological models, and, in combination with external information on $\Omegab$, have provided some of the earliest and most robust constraints on the cosmic matter density, $\Omegam$ \citep{White1993Natur.366..429W, David995ApJ...445..578D, White1995MNRAS.273...72W, Evrard9701148, Ettori9901304, Mohr9901281, Allen0205007, Ettori0211335}.
In addition, values of $\fgas$ inferred from cluster data are sensitive to the luminosity and angular-diameter distances between the observer and target; since the evolution of $\fgas$ in massive clusters is theoretically constrained to be minimal, such measurements also provide constraints on cosmological distances as a function of redshift, and thus on models of dark energy \citep{Sasaki9611033, Pen9610090}.
Over the past two decades, such data have consistently provided cosmological constraints comparable to those of other low-redshift probes, even with small samples of relaxed clusters (\citealt{Allen0205007, Allen0405340, Allen0706.0033, Ettori0211335, Ettori0904.2740}; \citealt*{Rapetti0409574}; \citealt{Mantz1402.6212}).
The \fgas{} approach complements tests based on the number density of clusters, and has independently provided similarly powerful constraints on dark energy \citep*{Allen1103.4829}.

Constraints on $\Omegam$ from the absolute value of \fgas{} have generally been systematically limited by uncertainty in the accuracy of cluster mass determinations \citep[][hereafter \amalmhkbers]{Applegate1509.02162}.
In contrast, dark energy constraints based on the apparent evolution of $\fgas$ have been limited by the redshift range of the available data, and the precision of measurements at high redshifts \citep[][hereafter \mamrakls]{Mantz1402.6212}.
While Sunyaev-Zel'dovich (SZ) surveys now routinely discover new clusters at redshifts $z>0.5$ (e.g., \citealt{Planck1303.5089, Planck1502.01598, Bleem1409.0850, Bleem1910.04121, Hilton1709.05600, Hilton2009.11043}), the restriction of the cosmological sample to the most dynamically relaxed systems (required to limit observational systematics), and the additional X-ray observations needed to identify clusters as relaxed and provide $\fgas$ measurements, means that relatively few new, high-redshift clusters have made their way into these studies.
In short, new relaxed clusters at high redshifts, or deeper observations of those already in the cosmological sample, can be expected to have an outsized impact on dark energy constraints from \fgas{}.
In addition, any extension of the sampled redshift range at the low-redshift end, where precise measurements can be obtained with comparatively short exposures, will disproportionately improve constraints from the method (\mamrakls).

In this work, we report the impact of improving and expanding the \fgas{} data set at both low and high redshifts.
At $z\gtsim0.6$, we incorporate significantly deeper data for several of the known relaxed clusters, compared with previous work, and add two systems, at $z=0.972$ and $z=1.160$, that have not previously been employed in this context.
At low redshifts, we incorporate a precise \fgas{} measurement for the Perseus Cluster (Abell~426), based on a new \Chandra{} mosaic, extending the sample from $z=0.078$ (Abell~2029) down to $z=0.018$.

This paper is structured as follows.
In Section~\ref{sec:data}, we describe the analysis of X-ray and weak gravitational lensing data employed in this work, with particular attention to the relatively challenging case of the X-ray observations of Perseus.
Section~\ref{sec:model} reviews the cosmological model fitted to the data, including allowances for various systematic uncertainties, while Section~\ref{sec:results} presents the resulting constraints on cosmological parameters from the $\fgas$ data alone, and in combination with other probes.
We conclude in Section~\ref{sec:conclusion}.
In general, quoted fitted parameter values refer to the modes of the corresponding marginalized posterior probability distributions, and quoted uncertainties refer to the 68.3 per cent probability highest posterior density (HPD) intervals.
In plots showing joint parameter constraints, dark and light shading respectively indicate the marginalized 68.3 and 95.4 per cent probability HPD regions.

\section{Data} \label{sec:data}

\subsection{X-ray data} \label{sec:xray}

The galaxy clusters used in this study are the most dynamically relaxed, hottest clusters known that have sufficiently deep \Chandra{} data to enable the requisite measurements.
Specifically, we require the clusters to have relaxed X-ray morphologies (as a proxy for true dynamical relaxation) according to the Symmetry-Peakiness-Alignment (SPA) criterion of \citet[][hereafter \mamslu]{Mantz1502.06020}.
The restriction to the most relaxed systems is intended to minimize systematic uncertainty and scatter associated with departures from hydrostatic equilibrium and spherical symmetry.
We additionally require that the ICM temperature in the isothermal part of the temperature profile be $\ge5$\,keV, to ensure that the selected clusters have genuinely deep gravitational potentials, such that their internal dynamics are gravitationally rather than astrophysically dominated at the radii of interest ($\sim r_{2500}$\footnote{This characteristic radius is defined such that the average enclosed density is 2500 times the critical density at the cluster's redshift.}).
This also reduces systematic uncertainties associated with cluster formation and active galactic nucleus (AGN) feedback compared with less massive systems.

A sample of 40 clusters meeting these criteria was constructed and employed for cosmological studies (\mamrakls, \mamslu).
To that sample, we add 4 clusters that had not been observed or did not have adequate \Chandra{} data at the time of the original search: the Perseus Cluster, RCS~J1447+0828, SPT~J0615$-$5746, and SPT~J2215$-$3537.
We also incorporate new observations of the original 40 clusters where available.
Compared with that of \mamrakls, the new data set extends to both lower and higher redshifts, and has significantly deeper exposures for targets at $0.6<z<0.9$, increasing its utility for constraining dark energy models.
The \Chandra{} data are summarized in Appendix~\ref{sec:data_table}; after cleaning, the total observing time is 4.9\,Ms (compared with 3.1\,Ms used by \mamrakls).

Our procedures for reducing and analyzing the \Chandra{} data, and extracting constraints on \fgas{} for each cluster, are unchanged from our previous work, and are described by \mamrakls{} and \mamslu{} (though note the special case of Perseus, addressed in the next section).
We do, however, use updated versions of the \Chandra{} analysis software and calibration files, namely {\sc ciao} 4.9 and {\sc caldb} 4.7.5.1.\footnote{Since the SPT~J2215$-$3537 data are more recent than these calibration files, we use {\sc ciao} 4.12 and {\sc caldb} 4.9.21 in that case.}
Among the updates is a retroactive change to the time-dependent model of the contaminant accumulating on the ACIS detectors.
We have not examined the impact of the calibration changes on individual clusters in detail, i.e.\ using only the old data reduced with the old and new calibrations, but note that in all cases our old and new constraints on gas density agree extremely well, such that systematic differences from \mamrakls{} in \fgas{} estimates can be attributed to changes in the measured temperatures. \footnote{The later entries in our previous series of papers on relaxed clusters (\citealt{Mantz1509.01322, Mantz1607.04686}; \amalmhkbers) employed an intermediate calibration version, which produces temperature profiles indistinguishable from those in this work.}
For redshifts $<0.6$, where the data set is best populated, the average impact of the calibration update is a marginal decrease in measured ICM temperatures, leading to a $\sim4$ per cent increase in \fgas{} estimates.
For each of the clusters at $0.6<z<1$, our new analysis incorporates a significant amount of new data, and thus these \fgas{} constraints have shrunk, while remaining consistent with the (significantly less precise) previous measurements.
For CL\,J1415.2+3612 and 3C\,186, with $1.0<z<1.1$, the updates reduce \fgas{} slightly while increasing its uncertainty, with the overall shift being small compared with the error bars.
We note that all these effects are within the scope of the systematic allowances employed in previous work and reprised here. 

\begin{figure}
  \centering
  \includegraphics{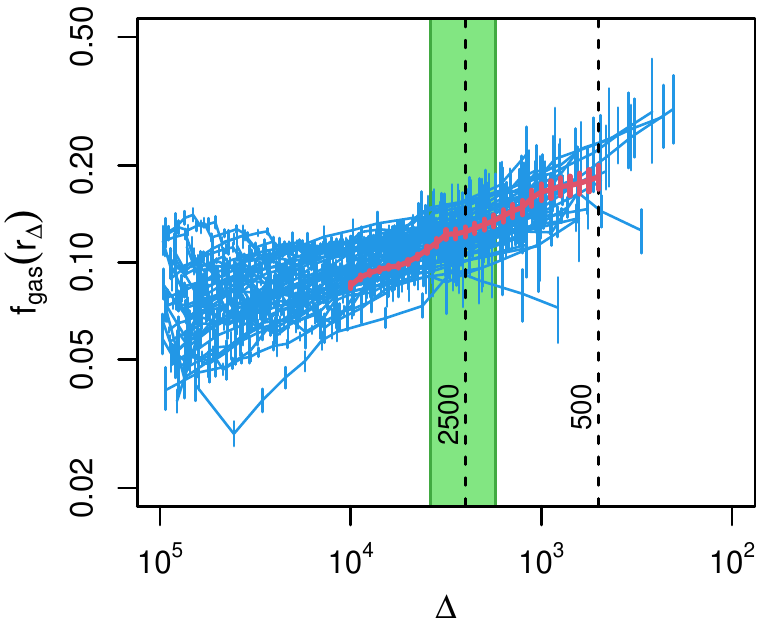}
  \caption{
    Differential \fgas{} profiles as a function of overdensity, calculated for our reference cosmology.
    The shaded region shows the typical range in $\Delta$ corresponding to the 0.8--$1.2\,r_{2500}$ shell where our cosmological measurements are made; by this point the dispersion is small compared with that seen at small radii (large overdensities).
    Results for the Perseus Cluster are shown with thicker, red lines (see Section~\ref{sec:perseus}).
  }
  \label{fig:fgas_profiles}
\end{figure}

The result of the X-ray analysis is simultaneous constraints on the gas mass and total mass profiles of each cluster, where the former is constructed non-parametrically and the latter assumes a parametrized \citet*[][hereafter \NFW]{Navarro9611107} form, as well as hydrostatic equilibrium.
Figure~\ref{fig:fgas_profiles} shows the resulting differential \fgas{} profiles (that is, the ratio of gas to total mass at a given radius, not interior to that radius) for the cluster sample.
The ordinate is the ``overdensity'', $\Delta$, defined as the ratio of the mean enclosed density at a given radius, $r_\Delta$, to the critical density of the Universe at the cluster's redshift.
For a monotonically decreasing density profile (such as the \NFW{} profile), $\Delta$ is thus a monotonically decreasing proxy for radius, for which the self-similar nature of many ICM thermodynamic profiles becomes clear (e.g., \citealt{Mantz1509.01322}).

\begin{figure*}
  \centering
  \includegraphics{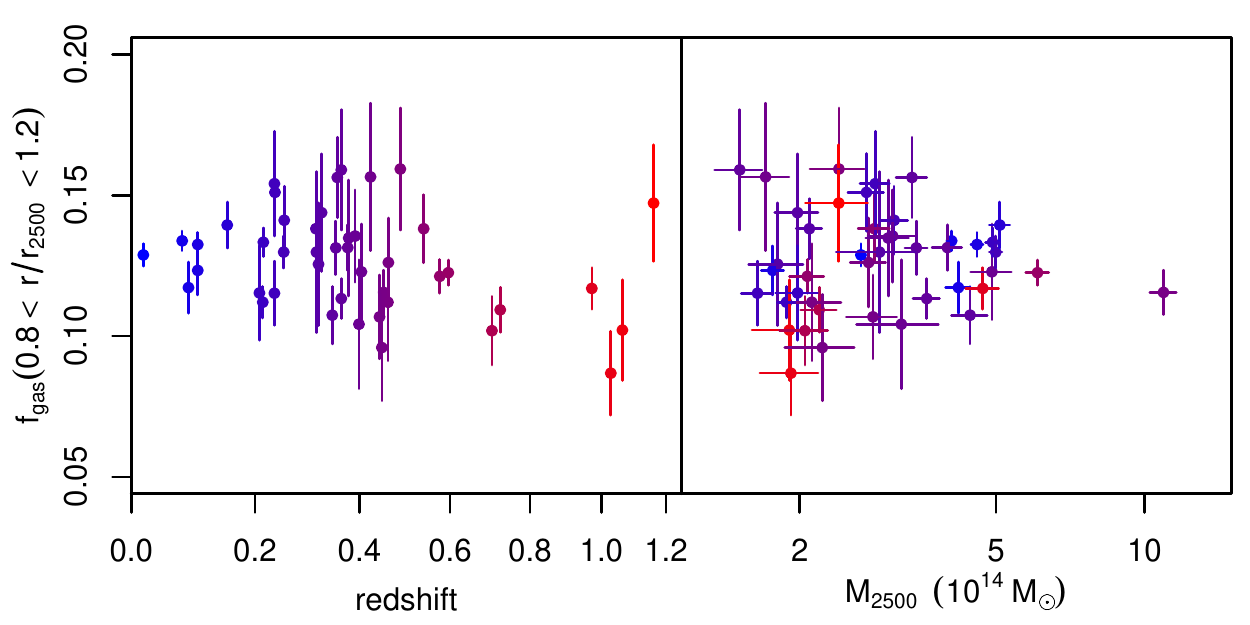}
  \caption{
    Gas mass fractions for our cluster sample, measured in our reference cosmology, as a function of redshift (left) and mass (right).
    Color coding is blue to red with increasing redshift.
    Note that the measurement uncertainties on $\fgas$ and $M_{2500}$ are typically strongly anti-correlated; this is not reflected in the way the error bars are displayed, but is accounted for in our analysis (Section~\ref{sec:model}).
  }
  \label{fig:fgas_zM}
\end{figure*}

As input to the cosmological likelihood, the X-ray measurements are summarized as constraints on the total mass within $r_{2500}$, $M_{2500}^\mathrm{ref}$, and the gas mass fraction in a shell spanning radii of 0.8--1.2\,$r_{2500}$, $\fgas^\mathrm{ref}$ (Table~\ref{tab:fgas_mass}; see discussion in \mamrakls{}).
The superscript ``ref'' indicates that these quantities are computed assuming a reference \LCDM{} model with $h=0.7$, $\Omegam=0.3$ and $\Omegal=0.7$, with the likelihood function accounting for this assumption (Section~\ref{sec:model}).
Figure~\ref{fig:fgas_zM} shows the measured (reference) gas mass fractions as a function of redshift and mass.

\begin{table*}
  \centering
  \caption{
    Redshifts, gas masses, total masses and gas-mass fractions of clusters in our sample from our X-ray analysis.
    Apart from redshift, these quantities are computed for our reference \LCDM{} cosmology; the applicable radial range is given in the header.
    Quoted error bars account for statistical uncertainties only. 
    Note that the total mass values listed here do not incorporate the calibration from weak lensing data, which is accounted for later in our analysis; a ``$\ast$'' indicates clusters for which we use lensing data (see \mamrakls{} and \amalmhkbers{}).
  }
  \vspace{1ex}
  \begin{tabular}{lcr@{ $\pm$ }lr@{ $\pm$ }lr@{ $\pm$ }lr@{ $\pm$ }l}
    \hline
    Cluster & $z$ & \multicolumn{2}{c}{$\Mgas^\mathrm{ref}$ $(10^{13}\Msun)$} & \multicolumn{2}{c}{$\Mtot^\mathrm{ref}$ $(10^{14}\Msun)$} & \multicolumn{2}{c}{$\fgas^\mathrm{ref}$} & \multicolumn{2}{c}{$M_{2500}^\mathrm{ref}$ $(10^{14}\Msun)$}\vspace{0.75ex}\\
    & & \multicolumn{2}{c}{0.8--1.2\,$r_{2500}$} & \multicolumn{2}{c}{0.8--1.2\,$r_{2500}$} & \multicolumn{2}{c}{0.8--1.2\,$r_{2500}$} & \multicolumn{2}{c}{0.0--1.0\,$r_{2500}$} \\
    \hline\vspace{-3ex}\\
    \input{tables/fgas_mass_table}
    \hline
  \end{tabular}
  \label{tab:fgas_mass}
\end{table*}

\subsection{Perseus Cluster} \label{sec:perseus}

Measurements of \fgas{} for the Perseus Cluster using \Chandra{} have been enabled by a mosaic of observations providing nearly complete coverage of the cluster out to $\sim1.2\,r_{2500}$ in radius, and limited azimuthal coverage in 8 directions out to $\sim r_{500}$ (proposal ID 16800086).
Perseus is unique in our sample in its large angular extent and the short duration of many of the individual pointings at the radii of interest (5\,ks), leading to a number of specializations of our standard analysis.

For these short observations, it is not possible to clean the data of point source emission to the level where the blank-sky fields normally used to model the extragalactic X-ray and particle backgrounds apply.
Furthermore, Perseus lies behind the Galactic plane, making the Galactic soft X-ray foreground significantly stronger in this field than in the blank-sky data.
To handle these issues, we constructed a background plus foreground model as follows.
The brightest point sources were identified and masked using preliminary results from the Cluster AGN Topography Survey pipeline (CATS; Canning et~al., in preparation).
To account for fainter sources, we include in the spectral model a power law component with index 1.4, whose normalization was computed based on the flux limit for detection by the CATS algorithm and the AGN luminosity function model of \citet{Miyaji1503.00056}.
This model normalization varies on an observation-by-observation basis, dependent on the exposure time and local cluster surface brightness;  
for the $\sim5$\,ks observations that predominate at radii $\sim r_{2500}$, a typical AGN detection flux limit is $\sim2.8\E{-14}$\,erg\,s$^{-1}$cm$^{-2}$ in the 0.5--8\,keV band.
To model the particle-induced background, we use data obtained while ACIS was in a stowed position, including these in the analysis in an identical manner to what is generally done with the blank-sky data (see \mamrakls).
The Galactic foreground was modeled as the sum of two thermal emission components with Solar metallicity, one unabsorbed and one absorbed.
The respective temperatures of 0.0974\,keV and 0.221\,keV were determined by fitting ROSAT All-Sky Survey spectra, extracted from an annulus spanning cluster radii of $2^\circ$--$4^\circ$ (i.e., beyond $1.5\,r_{200}$; \citealt{Simionescu1102.2429}), excluding regions to the W and NW that are contaminated by nearby structures.
Given the role of point-source finding in this procedure, and its reliance on an accurate model of the point spread function, we use data only from ACIS CCDs 0--3 and 7.
We note that the models for the soft foreground and the residual AGN emission, given the above treatment, are significantly fainter than both the cluster emission and the particle-induced background at the radii of interest.

Another consequence of Perseus' position behind the Galaxy and its angular size is that the equivalent absorbing hydrogen column density, $\NH$, is high, and varies across the $\gtsim 2^\circ$ diameter of the cluster.
In particular, \ion{H}{i} surveys reveal an overall gradient across the cluster.
However, since \NH{} values based on only \ion{H}{i} measurements are known to be inaccurate for sufficiently dense lines of sight ($>10^{21}\cm^{-2}$), we allow the overall column density to be variable in our analysis, taking only its spatial variation from the LAB \ion{H}{i} survey \citep{Kalberla0504140}.
We checked the adequacy of modeling only this overall shift in \NH{} by also performing independent fits in projection to spectra extracted over individual CCDs for every observation, each time modeling thermal emission from the cluster, the Galactic absorption, and the foreground and background components discussed above.
Those fits are consistent with a uniform, overall increase in \NH{} by a factor of $\sim1.3$ compared with the LAB values, apart from a statistically significant trend with radius at radii $<100$\,kpc (conservatively).
The latter likely reflects inadequacy of the single-temperature model for the cluster emission rather than the nature of the Galactic absorption, and data at these small radii are excluded from our final analysis based on other considerations, discussed below.
Our fitted values of \NH{} are consistent with the correction suggested by \citet{Willingale1303.0843}, based on comparing measurements of both \ion{H}{i} and molecular hydrogen to survey \ion{H}{i} values.

\begin{figure}
  \centering
  \includegraphics[scale=0.25]{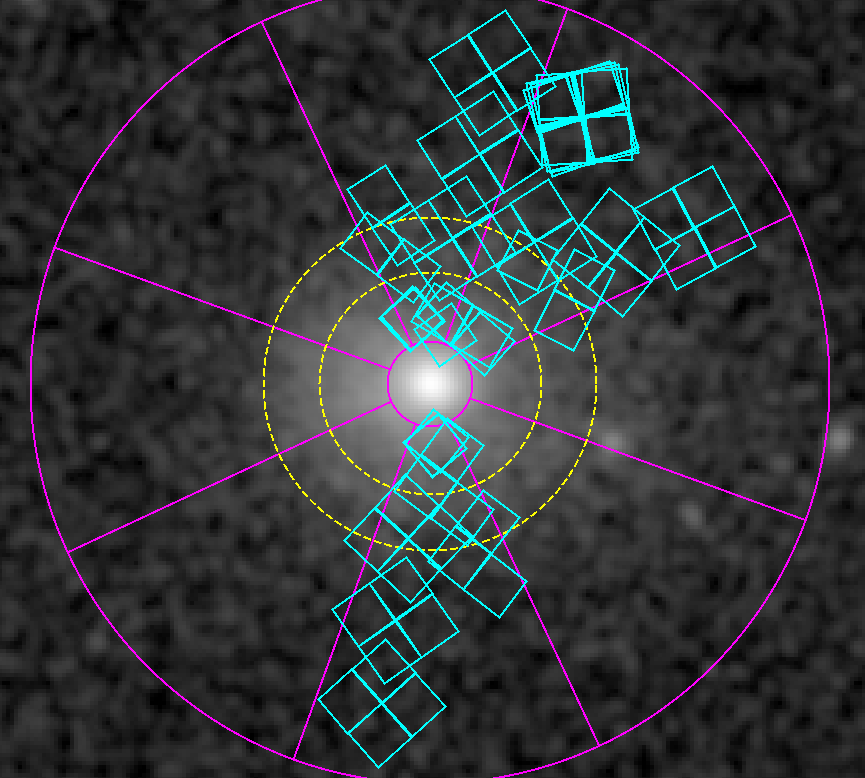}
  \caption{
    Smoothed ROSAT All-Sky Survey image of the Perseus Cluster \citep{Truemper1993Sci...260.1769T}, with per-CCD fields of view of the \Chandra{} data used in our analysis outlined in cyan (a subset of the available central pointings and wider mosaic).
    Magenta lines delimit the 8 sectors defined, of which we use 3, with inner and outer circles showing radii of $8'$ and $75'$.
    The dashed, yellow circles delineate the radial range 0.8--1.2\,$r_{2500}$, as estimated in Section~\ref{sec:perseus}.
  }
  \label{fig:perseus_image}
\end{figure}

While Perseus' global morphology satisfies the SPA criteria for inclusion in the \fgas{} sample, its emission is not perfectly circularly or elliptically symmetric.
Departures from symmetry exist at all radii, associated with large-scale sloshing of the gas, as well as a cold front aligned with the cluster's major axis \citep{Simionescu1208.2990}.
In this, Perseus is not necessarily different from any other cluster, including the most relaxed examples known.
However, its large X-ray flux, combined with our high spatial resolution, means that azimuthal variations are detected at extremely high significance, even in our shallow \Chandra{} data.
Consequently, our usual assumption that the ICM is characterized by a single density at each radius results in a poor fit that can bias the measurement of temperature, when a single emission model is used to describe the data at all azimuths.
Following \citet{Urban1307.3592}, we divide the cluster into 8 sectors (divided by position angles 25$^\circ$, 70$^\circ$, 115$^\circ$, \ldots{} E of N);
when fitting the data in these individual sectors at radii $>8'$, we find that the variations in brightness are small enough to obtain acceptable fits.
Note that this exclusion is larger than the $100$\,kpc scale discussed above (at this redshift, $1'\approx21.8$\,kpc).
Figure~\ref{fig:perseus_image} shows the spatial layout of the observations employed below.

For the measurement of \fgas{}, we are only interested in those sectors that lack large-scale cold fronts or signatures of sloshing at the radii of interest.
Again following \citet{Urban1307.3592}, we henceforth restrict the analysis to the N, NW and S sectors, as the NE, E and SE sectors intersect the largest cold front along the cluster's major axis (approximately E--W), and the W and SW sectors intersect a bright sloshing feature (effectively a minor cold front).
We constrain deprojected density, temperature and mass profiles for the N, NW and S sectors independently, using our standard methodology (Section~\ref{sec:xray}; note that this entails independent solutions for both the gas density and the gravitational potential in each sector), with the modified treatment of Galactic absorption and background and foreground components described above, and considering only data at radii $>8'$ from the cluster center. In addition, the \Chandra{} coverage of the NW arm has a gap at radii of approximately $18.5'$--$21.5'$ (405--470\,kpc), a range within which our model nominally has 3 free temperatures.
In order to perform the deprojection to the NW, we marginalize each of these temperatures over a uniform prior spanning 7--9\,keV, centered on the best fit of $\approx8$\,keV found in the N sector; given the use of a parametrized mass model and the assumption of hydrostatic equilibrium, an additional prior on the density profile is not necessary.

\begin{figure}
  \centering
  \includegraphics{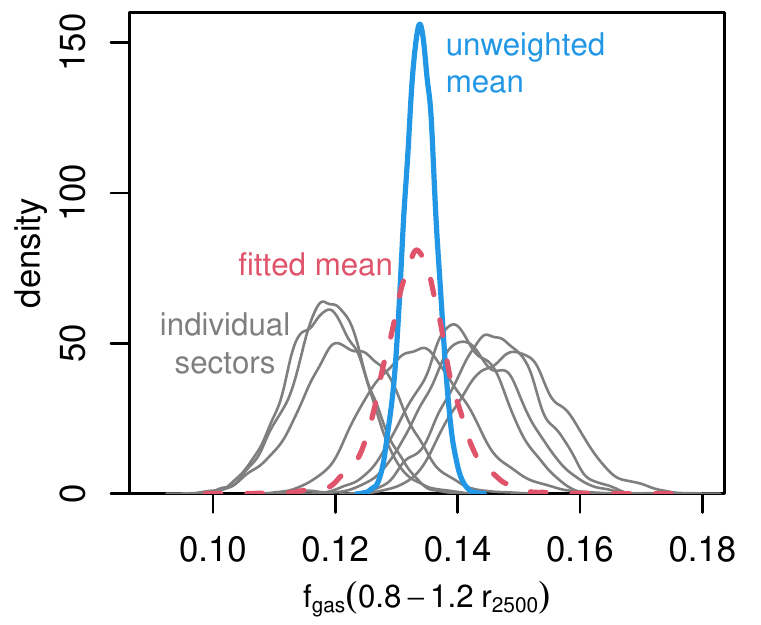}
  \caption{
    Posterior distributions of \fgas{} measured independently from data in 8 sectors about the center of Abell~2029 (gray, solid curves).
    The solid, blue curve shows the distribution of the unweighted mean of \fgas{} in the 8 sectors, marginalizing over their individual posteriors.
    The red, dashed curve shows the constraint on the mean \fgas{} in the 8 sectors when a Gaussian intrinsic scatter is fitted simultaneously.
  }
  \label{fig:a2029sec}
\end{figure}

We are then left with the question of how to combine 3 measurements of the \fgas{} profile along different directions into a single estimate for Perseus.
Scatter among estimates made in this way is expected, driven primarily by asphericity and the resulting ellipticity of the cluster emission in the plane of the sky (e.g.\ \citealt{Roncarelli1303.6506, Ansarifard1911.07878}).
To gain some insight, we turned to the single other cluster in the sample for which an azimuthally resolved analysis is possible, Abell~2029.
In this case, there are no visible substructures outside the cluster core that would discourage the use of particular regions within the cluster.
We therefore divided the data into 8 sectors, oriented sensibly with respect to the cluster's major and minor axes, and measured deprojected \fgas{} profiles independently using the data in each sector, as was done in Perseus.
The posterior distributions for \fgas{} in the 0.8--1.2\,$r_{2500}$ shell, as determined from each sector, are shown in Figure~\ref{fig:a2029sec}.
The observed variation is primarily due to differences in the derived gas mass profiles, with the total mass profiles (and thus $r_{2500}$) being more mutually consistent.

We considered two methods of combining the information from each arm: fitting a mean plus Gaussian intrinsic scatter, and computing the unweighted mean while marginalizing over the individual posteriors from each sector.
The former method is more interesting in that it can provide a measurement of the scatter that we expect to be present due to asphericity, given sufficient data.
The second has the advantage of providing an estimate of the mean even when there is insufficient data to simultaneously constrain the scatter, as is the case for the 3 sectors of Perseus.
For Abell~2029, we find that these methods produce very similar values of the mean \fgas{}, albeit with different uncertainties, as can be seen in Figure~\ref{fig:a2029sec}.
From the mean+scatter fit, the intrinsic scatter is constrained to be $8\pm4$ per cent, consistent with predictions for relaxed clusters \citep{Roncarelli1303.6506, Ansarifard1911.07878}.

In the case of Perseus, the same mean+scatter model cannot be constrained based on measurements from only 3 sectors.
We can, however, fit a mean \fgas{} value when fixing the intrinsic scatter to the best-fitting value from the Abell~2029 analysis (i.e., assuming the sector-to-sector scatter is similar).  
This turns out to produce essentially identical results to the unweighted-mean method applied to Perseus, including the uncertainty estimate (unlike in Abell~2029).
We adopt the simpler unweighted-mean method for our final Perseus results, arriving at the estimate of $\fgas^\mathrm{ref} = 0.129\pm0.004$.
The differential \fgas{} profile, computed in the same way as a function of overdensity, is shown in Figure~\ref{fig:fgas_profiles}.
Applying the same procedure to the estimate of $r_{2500}$, we find $r_{2500}=569\pm3$\,kpc ($26.05'\pm0.13'$), or, equivalently, $M_{2500}=(2.66\pm0.04)\E{14}\Msun$.

Systematic azimuthal variations in ICM density at a given radius cause an overestimate of the density when the data at all azimuths are straightforwardly combined in a single fit, rather than independent fits in multiple sectors, as above.
In the simplest approximation, lognormal variations in the true density of standard deviation $\sigma$ lead to a multiplicative bias of $1+\sigma$ in the estimated density.
This is compatible with our analysis of Abell~2029, where the best-fitting \fgas{} value from an azimuthally averaged analysis exceeds the mean \fgas{} among 8 sectors by 5 per cent, consistent with the $8\pm4$ per cent scatter found above.
To the extent that this scatter is driven by the elliptical shape of the gas, one might consider correcting such a bias on a cluster-by-cluster basis based on their projected ellipticities at $r_{2500}$.
However, we note that the \fgas{} variations among the sectors in Figure~\ref{fig:a2029sec} do not quite correspond to each sector's placement with respect to the major and minor axes, as one would expect in this simplistic interpretation.
Furthermore, the exposure time as a function of azimuth for Abell~2029 is much more heterogenous than for other clusters in the sample, with significantly deeper observations along the major axis; it is possible that this may introduce additional scatter that does not generalize to other cluster observations.
We therefore choose not to attempt a general correction for this bias based on the single cluster where we can obtain suitable measurements in several sectors, though further investigation in future work is certainly warranted.
Instead, we adopt the improved \fgas{} estimates based on azimuthally resolved measurements for Perseus and Abell~2029 only.
We expect any small shift in their \fgas{} values with respect to other clusters in the sample so introduced to be subsumed into the global intrinsic scatter parameter in our cosmological analysis.
Moreover, the relatively small values of this intrinsic scatter that we find a posteriori imply that such an effect cannot be large.

\subsection{Weak gravitational lensing}

To constrain any overall bias of the masses estimated from X-ray data, we incorporate weak gravitational lensing measurements from the Weighing the Giants project \citep{von-der-Linden1208.0597, Kelly1208.0602, Applegate1208.0605}, as in \mamrakls.
Specifically, we use the subset of lensing measurements for relaxed clusters in our sample.
There are 12 of these, of which 6 have 5-band data sufficient to obtain robust photometric redshifts for individual source galaxies, and 6 for which ``color-cut'' methods have been employed instead; for consistency, we use the color-cut estimates for all 12 clusters, accounting for the additional systematic uncertainty as required \citep{Applegate1208.0605}.
While the data have not changed from previous work, we do incorporate updated systematics modeling from \amalmhkbers, in particular a potential redshift dependence in the absolute accuracy of masses estimated using these methods (see Section~\ref{sec:model}).

\section{Model} \label{sec:model}

Our modeling of the data follows previous work (\mamrakls; \amalmhkbers), with the exception that here we allow \fgas{} to potentially vary with mass as a power law, constraining its slope, $\alpha$, simultaneously with other parameters.
We describe the model and its parameters below, and refer the reader to the works cited above for complete details.

We model the true gas mass fraction in a spherical shell spanning radii of  0.8--1.2\,$r_{2500}$ as
\begin{equation} \label{eq:fgastrue}
  \fgas(z,M_{2500}) = \Upsilon(z,M_{2500}) \frac{\Omegab}{\Omegam},
\end{equation}
where $\Upsilon(z, M_{2500})$ is the gas depletion of massive clusters, parameterized as
\begin{equation} \label{eq:depletion}
  \Upsilon(z, M_{2500}) = \Upsilon_0 (1 + \Upsilon_1 z) \left( \frac{M_{2500}}{3\E{14}\Msun} \right)^\alpha.
\end{equation}
The model includes a log-normal intrinsic scatter about $\fgas(z,M_{2500})$, with standard deviation $\sigma_f$. 

\mamrakls{} adopted a uniform prior on $\Upsilon_0$ centered on the average prediction of the hydrodynamic simulations of \citet{Battaglia1209.4082} and \citet{Planelles1209.5058}, which include radiative cooling, star formation, and heating from AGN and supernova feedback.
However, a misinterpretation of one of the results of \citet{Planelles1209.5058} led this central value to be somewhat larger than it should have been.
We correct this here, centering the prior at $\Upsilon_0=0.79$, maintaining the full width of 20 per cent (2.3 times the difference between the two simulations) from \mamrakls.
We maintain the uniform prior on $\Upsilon_1$ between $-0.05$ and $+0.05$ from \mamrakls{}, which is not impacted by the issue noted above.

Even in the absence of statistical uncertainties, systematic biases and the assumption of a reference cosmology may cause the measured gas mass fractions to  differ from $\fgas(z,M_{2500})$:
\begin{equation} \label{eq:fgasref}
  \fgas^\mathrm{ref} = K(z) \, A(z)^{\eta_f} \left[ \frac{d^\mathrm{ref}(z)}{d(z)} \right]^{3/2} \fgas(z,M_{2500}).
\end{equation}
Here the ratio $[d^\mathrm{ref}(z)/d(z)]^{3/2}$ accounts for the impact of the assumed reference cosmology on \fgas{} measurements within a fixed angular aperture, while the term
\begin{equation} \label{eq:thetarat}
  A(z) = \frac{\theta^\mathrm{ref}_{2500}}{\theta_{2500}} = \frac{H(z) \, d(z)}{\left[H(z) \, d(z)\right]^\mathrm{ref}}
\end{equation}
accounts for the relatively smaller correction that arises from the dependence of the measurement aperture itself on the reference model.
We empirically measure a power-law slope of the aperature-measured \fgas{} with radius of $\eta_f=0.390\pm0.024$; note that this is not the slope of $\fgas(r)$, but of the gas mass fraction integrated in a shell, $\fgas(0.8x<r<1.2x)$, as $x$ varies in the neighborhood of $r_{2500}$.
The final term,
\begin{equation}
  K(z) = K_0 (1 + K_1 z),
\end{equation}
parametrizes a potentially redshift-dependent bias in the \fgas{} measurements due to bias in the total mass estimates.
The $K_0$ parameter is well constrained by our weak lensing data (see below), while we marginalize $K_1$ over the range $-0.05<K_1<0.05$.

Biases in the X-ray gas mass estimates may also be present, in particular due to azimuthal variations in gas density at a given circular radius, which leads the root-mean-square density measured from the data to exceed the true mean density.
At radii $\sim r_{2500}$ in relaxed clusters, simulations place the bias in density (and thus gas mass) at $\ltsim5$ per cent \citep{Battaglia1209.4082, Battaglia1405.3346, Roncarelli1303.6506, Planelles1612.07260, Ansarifard1911.07878}. 
Observations are in broad agreement with this limit \citep{Eckert1310.8389, Zhuravleva1906.06346}, but do not yet address the specific cluster selection and radial range of interest here.
Lacking such direct, empirical input, and given that any bias is expected to be subdominant to the systematic uncertainty in the mass measurements, we do not explicitly account for potential bias in gas masses in this work.
We note, however, that a positive bias of 5 per cent would straightforwardly impact constraints from the absolute value of \fgas{} (Section~\ref{sec:lowz}) by the same amount.
For $\Omegam$, for example, this would result in a shift of 0.01, or one quarter of our posterior uncertainty.

Analogously to \fgas{}, we can write the relationship between true mass and mass as estimated for a reference cosmological model as
\begin{equation} \label{eq:m2500ref}
  M_{2500}^\mathrm{ref} = A(z)^{\eta_M} \left[ \frac{d^\mathrm{ref}(z)}{d(z)} \right] M_{2500},
\end{equation}
with the slope of the mass profile near $r_{2500}$, $\eta_m=1.065\pm0.016$, measured empirically from the X-ray fits.

The joint posterior distributions of $\fgas^\mathrm{ref}$ and $M_{2500}^\mathrm{ref}$ from the X-ray analysis are well described by bivariate log-normal distributions, and we model them as such, including their (generally substantial) anti-correlation.

Since the measured gravitational lensing shear depends on distances between the observer, lens and numerous background objects (as opposed to only the observer and cluster, as in the X-ray case), the cosmological dependence of inferred masses cannot be accounted for with simple factors of $d(z)/d^\mathrm{ref}(z)$, as above.
Instead, following \amalmhkbers, we directly incorporate the measured shear profiles and their Gaussian measurement uncertainties in our cosmological analysis.
We assume a log-normal distribution of lensing/X-ray mass ratios at $r_{2500}^\mathrm{ref}$, with mean $\ln K(z)$ and scatter $\sigma_K$, and marginalize over the mass constraints from the lensing and X-ray data, assuming an \NFW{} form of the mass profile in both cases.
Note that, while the same parametrized model is fit to both types of data, the constraints from each method are independent of the other.
In particular, the X-ray data are sufficient to constrain both parameters of the \NFW{} model in all cases, while for the lensing analysis we adopt a prior on the distribution of \NFW{} concentration parameters of massive, relaxed clusters motivated by simulations \citep{Neto0706.2919}.
The systematic error budget for the lensing mass estimates is discussed by \citet{Applegate1208.0605} and \amalmhkbers, and is described by a redshift-dependent bias of $W(z) = W_0 + W_1(z-0.4)$, with $W_0=0.96\pm0.09$ and $W_1=-0.09\pm0.03$.

Table~\ref{tab:fgas_params} summarizes the model parameters specific to the \fgas{} cosmological analysis, and the corresponding priors (see further discussion in \mamrakls).

\begin{figure}
  \centering
  \includegraphics{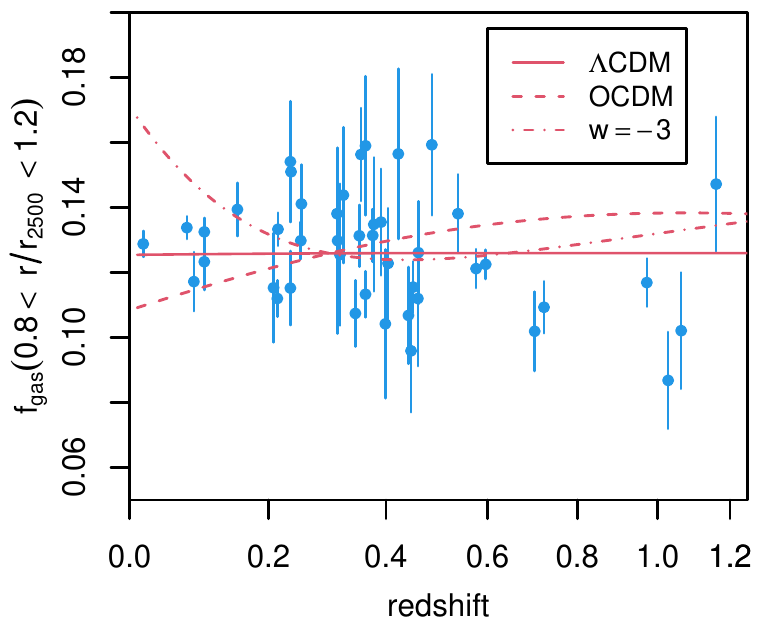}
  \caption{
    The \fgas{} data as a function of redshift, as in Figure~\ref{fig:fgas_zM}, are compared with predictions of 3 dark energy models.
    These predictions incorporate the complete model of Section~\ref{sec:model}, meaning that they are predictions for what \fgas{} values we would measure given the adopted reference cosmology, for nominal values for the nuisance parameters.
    The solid line shows predictions for a flat \LCDM{} model ($\Omegam=0.3$, $\Omegal=0.7$, $w=-1$; identical to the reference), the dashed line an open model ($\Omegam=0.3$, $\Omegal=0.0$), and the dot-dashed line a flat, constant-$w$ model ($\Omegam=0.3$, $\Omegade=0.7$, $w=-3$).
    To illustrate the different shapes of the curves as a function of redshift, they are normalized to intersect at $z=0.3$, which is approximately the weighted mean redshift of the data (note that this is a different and arguably better motivated choice than in the equivalent figure in \mamrakls{}).
    The figure thus demonstrates the redshift-dependent signal available to the $\fgas(z)$ data once the $\Omegam$ constraint from the normalization of $\fgas$ is accounted for, in particular emphasizing the role of data at the lowest redshifts in ruling out the more extreme models.
  }
  \label{fig:modfgasz}
\end{figure}

To illustrate the redshift dependence of the model, we show in Figure~\ref{fig:modfgasz} predictions in data space for 3 different cosmologies.
The curves are normalized to intersect at the weighted-mean redshift of the data, to better illustrate their different shapes as a function of redshift.
In this way, we can see how precise measurements across a wide range in redshifts, especially when extending to $z\approx0$, can place constraints on dark energy parameters.

\section{Results} \label{sec:results}

The posterior probability encoded by the model of Section~\ref{sec:model} is implemented as a stand-alone library.\footnote{\url{https://github.com/abmantz/fgas-cosmo}}
The results presented below were produced by including it in {\sc cosmomc}\footnote{\url{http://cosmologist.info/cosmomc/}} (\citealt{Lewis0205436}; version May 2020), with cosmological calculations provided by {\sc camb} \citep*{Lewis9911177}.
When analyzing the \fgas{} data alone, or in combination with simple, external priors, we use {\sc cosmomc}'s ``astro'' cosmological parametrization, with the free parameters and priors given by Table~\ref{tab:astro_params}.

We also compare and combine our results with those of other cosmological probes whose likelihoods are available in {\sc cosmomc}.
In particular, these include the final (2018) \Planck{} CMB temperature, polarization and lensing power spectra \citep{Planck1907.12875, Planck1807.06210},
type Ia supernova (SN) from the Pantheon project \citep{Scolnic1710.00845},
and Baryon Acoustic Oscillations (BAO).
In the case of BAO, we will show the final results from the Sloan Digital Sky Survey (SDSS) IV extended Baryon Oscillation Spectroscopic Survey (BOSS) \citep{eBOSS2007.08991};
however, as the data underlying those results is not publicly available at the time of our analysis, we use earlier data from the 6-degree Field Galaxy Survey ($z=0.106$; \citealt{Beutler1106.3366}), the SDSS DR7 main galaxy sample ($z=0.15$; \citealt{Ross1409.3242}) and the SDSS-III BOSS ($z=0.38$, 0.51 and 0.61; \citealt{Alam1607.03155}) when obtaining joint constraints.
When obtaining constraints using CMB data (alone or in combination with others), we use the ``theta'' cosmological parametrization in {\sc cosmomc}, with the free parameters and priors shown in Table~\ref{tab:theta_params}.
Though we do not perform a combined analysis with them, we also reproduce constraints from the Dark Energy Survey (DES) Year-3 analysis of galaxy clustering and weak lensing (``$3\times2$pt''; \citealt{DES2105.13549}) as a point of comparison.

In models with $w$ free, constraints from only the CMB data employed here are effectively unable to place an upper limit on the Hubble parameter, $H_0$ (equivalently, $h=H_0/100$\,km\,s$^{-1}$Mpc$^{-1}$).
The resulting contours (Figures~\ref{fig:lowz} and \ref{fig:constw}) may appear surprisingly unconstrained to some readers, due to the degeneracy between $h$ and other parameters, and the fact that we marginalize over a wider uniform prior on $h$ than is typical in the literature (0.2--2 rather than 0.4--1; e.g. \citealt{Planck1807.06209}).
While one could argue unconvincingly about which range better reflects our prior knowledge of $h$, we prefer not to show contours that appear to provide constraints on other parameters (e.g.\ \Omegam{}) that are, ultimately, due to the limited prior range of $h$.
The reader should interpret these particular contours as being indicative of which combinations of parameters are well constrained and which are degenerate, and thus how combinations of independent probes might improve constraints, without taking the absolute probability levels too seriously.

\subsection{Constraints from low-redshift \fgas{} data} \label{sec:lowz}

The dependence of the \fgas{} observable on the cosmological parameters, apart from the relatively minor $A(z)$ term, is $\fgas^\mathrm{ref} \propto d(z)^{-3/2} \Omegab/\Omegam$.
Thus, constraints can in principle be obtained from both the normalization of $\fgas^\mathrm{ref}(z)$ and its behavior with redshift.
The former depends the cosmological parameter combination $h^{3/2} \Omegab/\Omegam$, and can be measured to high statistical precision using only the low-redshift clusters in our sample.
We can, therefore, obtain constraints on this degenerate combination of parameters from a subset of the \fgas{} data.
The advantage of doing so is that these constraints are insensitive to the particular model of dark energy used, provided that its equation of state does not evolve strongly over the redshift range of the data employed.
Following \mamrakls{}, we obtained such constraints from the 6 clusters in our sample with $z<0.16$, marginalizing over a non-flat cosmological model with $w$ free, finding $h^{3/2} \Omegab/\Omegam = 0.096\pm0.013$. Since we do not use weak lensing data for any of these low-$z$ clusters, we employ  a Gaussian prior, $K_0=0.93\pm0.11$, obtained from our full analysis (Section~\ref{sec:misc}).
As mentioned in Section~\ref{sec:model}, this simplified approach neglects any correlations between cosmological parameters and the measured X-ray/lensing mass ratio, but we note that these are small compared with current uncertainties (see \amalmhkbers).
At present, the constraint on $h^{3/2} \Omegab/\Omegam$ is limited by this uncertainty in $K_0$, which is primarily due to the small number of clusters for which we use lensing data to calibrate our X-ray masses.

\begin{figure}
  \centering
  \includegraphics{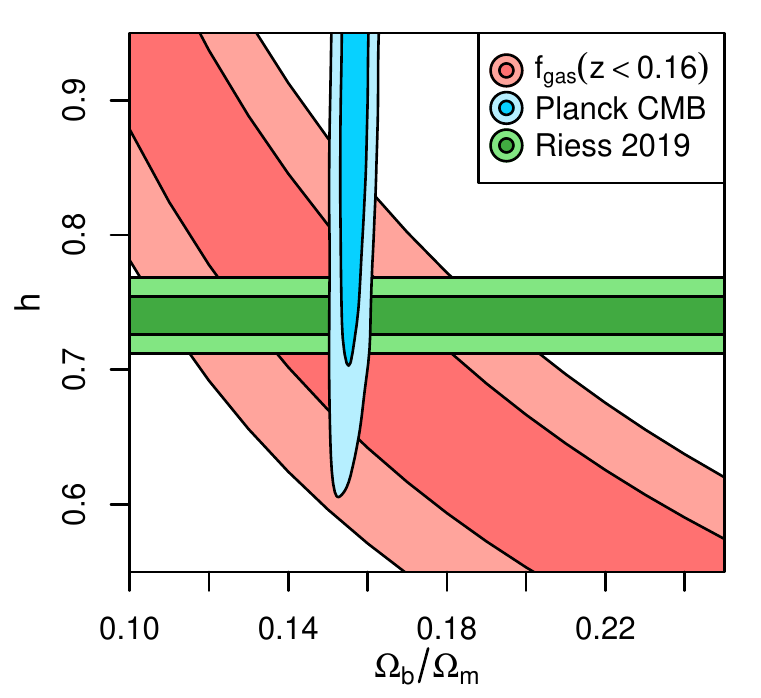}
  \caption{
    Constraints on the Hubble parameter and the cosmic baryon fraction from our \fgas{} data at $z<0.16$ only (red), \Planck{} CMB data \citep{Planck1907.12875, Planck1807.06210}, and the Cepheid distance ladder \citep{Riess1903.07603}.
    The \Planck{} analysis assumes a flat, constant-$w$ model, while the \fgas{} and distance-ladder constraints are essentially independent of the cosmological model.
  }
  \label{fig:lowz}
\end{figure}

The above constraint was obtained without external cosmological priors, but, as Figure~\ref{fig:lowz} illustrates, there are interesting complementarities with independent data.
We investigate three forms of such external information, with the results shown in Table~\ref{tab:lowz}.

\begin{table}
  \begin{center}
    \caption{
      Marginalized best-fitting values and 68.3 per cent probability credible intervals on cosmological parameters from our analysis of low-redshift ($z < 0.16$) cluster \fgas{} data, including systematic uncertainties.
      At the quoted precision, these constraints are identical for all cosmological models considered in this work.
      Columns 1--3 indicate whether priors on each cosmological quantity were included (see text; F01=\citealt{Freedman0012376}; P18=\citealt{Planck1807.06209}, R19=\citealt{Riess1903.07603}, H20=\citealt{Hsyu2005.12290}).
    }
    \label{tab:lowz}
    \vspace{1ex}
    \begin{tabular}{cccr@{ $=$ }r@{ $\pm$ }l}
      \hline
      \multicolumn{3}{c}{Prior} & \multicolumn{3}{c}{Constraint} \\
      $h$ & $\Omegab h^2$ & $\Omegab/\Omegam$ & \multicolumn{3}{c}{} \\
      \hline\vspace{-1.5ex}\\
      \input{tables/lowz_constraints}
      \hline
    \end{tabular}
  \end{center}
\end{table}

Combining our data with a prior on the cosmic baryon fraction, $\Omegab/\Omegam=0.156 \pm 0.003$, from {\it Planck}, we constrain the Hubble parameter to be $h=0.722\pm0.067$.
Note that, whereas the direct constraints placed on $h$ by the CMB data are precise only when assuming a flat \LCDM{} model, the CMB baryon fraction constraints hold much more widely (Figure~\ref{fig:lowz} shows contours for a constant-$w$ model, displaying essentially no correlation in the $h$ vs $\Omegab/\Omegam$ plane).
As the \fgas{} constraints considered here are similarly insensitive to the dark energy model assumed, the combined \fgas{}+CMB constraint on $h$ is an interesting one to compare to independent probes.
In particular, we prefer a value closer to that determined from the Cepheid-based distance ladder \citep{Riess1903.07603} than to the constraints from {\it Planck} when assuming flat \LCDM{}, although our results are consistent with both within uncertainties.
Forthcoming weak lensing data (Baumont et~al., in preparation; Wright et~al., in preparation) will roughly halve the uncertainty in $K_0$, translating directly to tighter constraints on $h$ from the \fgas+CMB combination.
We note that compatible results to those above were obtained from the combination of WMAP CMB data with \fgas{} data from \citet{Allen0706.0033} or \mamrakls{}, and the combination of BAO data with the \mamrakls{} \fgas{} data \citep{Holanda2006.06712}. 
Our contraints are also compatible with those obtained by enforcing consistency between X-ray and SZ measurements of the same ICM; the latter method has the benefit of not requiring even the limited external priors adopted here, but is significantly less precise at present (see \citealt{Wan2101.09389} for a recent discussion).

Conversely, adopting external priors on $h$ allows us to test for consistency with the baryon fraction measured from the CMB.
We find good agreement, within uncertainties, using priors from the Hubble Key project ($h=0.72\pm0.08$; \citealt{Freedman0012376}), {\it Planck} (assuming flat \LCDM{}; $h=0.674\pm0.005$; \citealt{Planck1807.06209}) or Cepheids ($h=0.7403\pm0.0142$; \citealt{Riess1903.07603}).

Combining the \fgas{} data with a prior on the baryon density, $\Omegab h^2=0.0215 \pm 0.0005$, based on estimates of the primordial helium and deuterium abundances \citep{Hsyu2005.12290}, provides the constraint $\Omegam h^{1/2} = 0.221 \pm 0.031$.
The weak residual dependence on $h$ means that we obtain nearly identical constraints on $\Omegam$ when additionally combining with any of the priors on $h$ listed above.
Using the results of \citet{Freedman0012376}, which are comfortably consistent with the more recent estimates, the low-redshift \fgas{} data yield $\Omegam=0.260 \pm 0.040$.

\subsection{$\Omegam$ from the evolution of $\fgas$} \label{sec:om}

\begin{figure}
  \centering
  \includegraphics{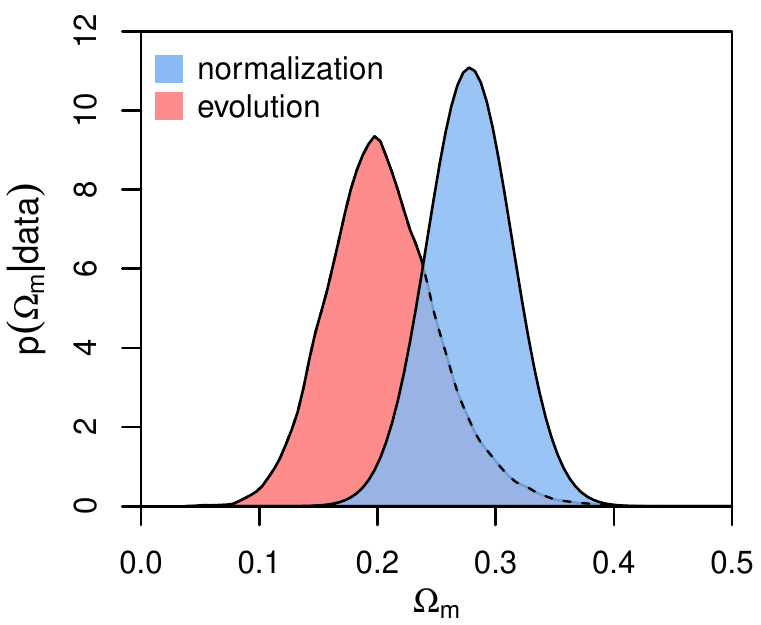}
  \caption{
    Posterior distributions for \Omegam{} based on \fgas{} data.
    The blue-shaded constraints are from the normalization of $\fgas(z)$ measured from only the $z<0.16$ clusters in our sample, using standard priors on $h$, $\Omegab h^2$ and the relevant nuisance parameters (Section~\ref{sec:lowz}).
    These results are insensitive to the assumed model of dark energy.
    The red-shaded results are from the shape of the measured $\fgas(z)$ curve, using our full sample without the priors that would produce a constraint from the normalization, and assuming a flat \LCDM{} model (Section~\ref{sec:om}).
  }
  \label{fig:omegam}
\end{figure}

While the normalization of $\fgas^\mathrm{ref}(z)$ constrains the combination $h^{3/2}\Omegab/\Omegam$, its behavior with redshift is sensitive to cosmology through the shape of $d(z)$.
Uniquely, $\Omegam$ contributes to the observed signal in both places.
For the flat \LCDM{} model, we can obtain precise constraints on \Omegam{} from the evolution of $\fgas^\mathrm{ref}(z)$ alone, and compare them to those from its normalization.
This can be accomplished in practice by widening any of the informative priors impacting the normalization of $\fgas(z)$ (i.e.\ on $h$, $\Omegab h^2$ or $\Ugas{}_{,0}$) until it no longer has an effect on the posterior distribution.
Our constraints from this procedure are compared with those of the preceding section in Figure~\ref{fig:omegam}; we find $\Omegam=0.200\pm0.044$, consistent with but marginally lower than the results from the normalization.

\subsection{Constraints from the full \fgas{} data set} \label{sec:allfgas}

\begin{figure*}
  \centering
  \includegraphics{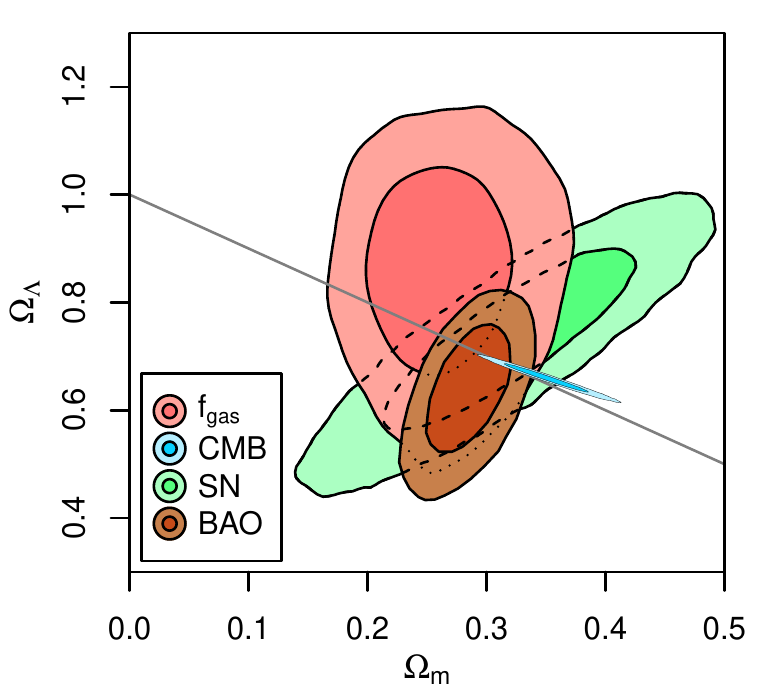}
  \hspace{10mm}
  \includegraphics{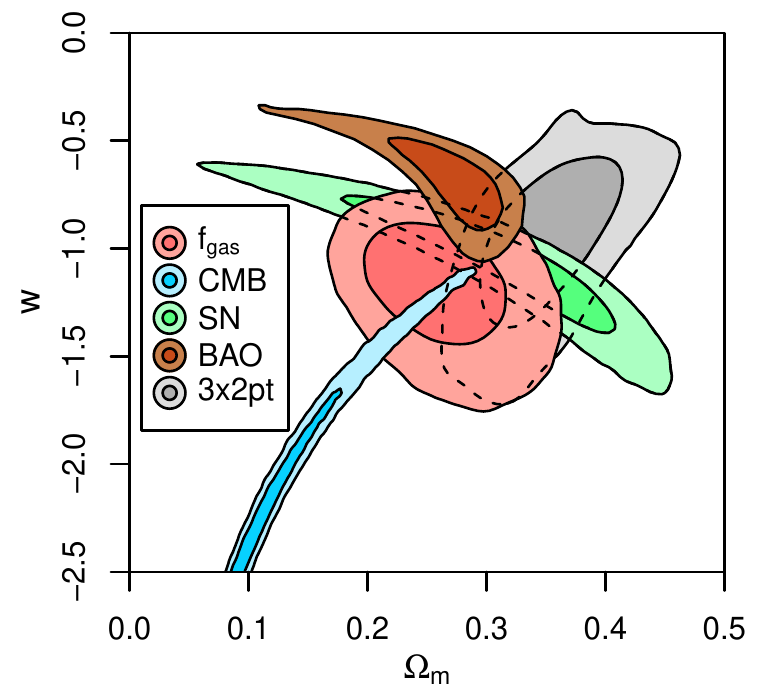}
  \caption{
    Constraints on parameters of the \LCDM{} (left) and flat, constant-$w$ (right) models from \fgas{} (this work), \Planck{} CMB \citep{Planck1907.12875, Planck1807.06210}, Pantheon SN \citep{Scolnic1710.00845}, BAO \citep{eBOSS2007.08991}, and galaxy clustering and lensing (``$3\times2$pt''; \citealt{DES2105.13549}) data. In the left panel, the solid, gray line indicates spatial flatness ($\Omegal=1-\Omegam$).
  }
  \label{fig:lcdm}
  \label{fig:constw}
\end{figure*}

We next report the constraints available from the combination of the complete \fgas{} data set with an external prior on $\Omegab h^2$ from \citet{Hsyu2005.12290}, and a broad prior on $h$ from \citet{Freedman0012376}.
Figure~\ref{fig:lcdm} shows our results as red-shaded contours for the non-flat \LCDM{} model (left) and flat constant-$w$ model (right).
The degeneracies of $\Omegal$ and $w$ with $\Omegam$ are modest, due to the direct constraint on $\Omegam$ coming from the normalization of $\fgas(z)$.
For the \LCDM{} model, we find $\Omegam=0.257\pm0.039$ and $\Omegal=0.865\pm0.119$, while for the flat, constant-$w$ model we find $\Omegam=0.256\pm0.037$ and $w=-1.13^{+0.17}_{-0.20}$ (Table~\ref{tab:fgas_constraints}).
Compared with \mamrakls, these constraints represent improvements of 41 per cent in the constraints on $\Omegal$ and 29 per cent on $w$ in the respective models.

\begin{table*}
  \begin{center}
    \caption{
      Marginalized best-fitting values and 68.3 per cent probability credible intervals on cosmological parameters of \LCDM{} and flat, constant-$w$ models from our data, in combination with priors on $h$ \citep{Freedman0012376} and $\Omegab h^2$ \citep{Hsyu2005.12290}.
    }
    \label{tab:fgas_constraints}
    \vspace{1ex}
    \begin{tabular}{ccccc}
      \hline
      Model & \Omegam{} & \Omegade{} & \phmin$\Omega_k$ & $w_0$ \\
      \hline\vspace{-1.5ex}\\
      \input{tables/fgas_constraints}
      \hline
    \end{tabular}
  \end{center}
\end{table*}

\begin{figure*}
  \centering
  \includegraphics{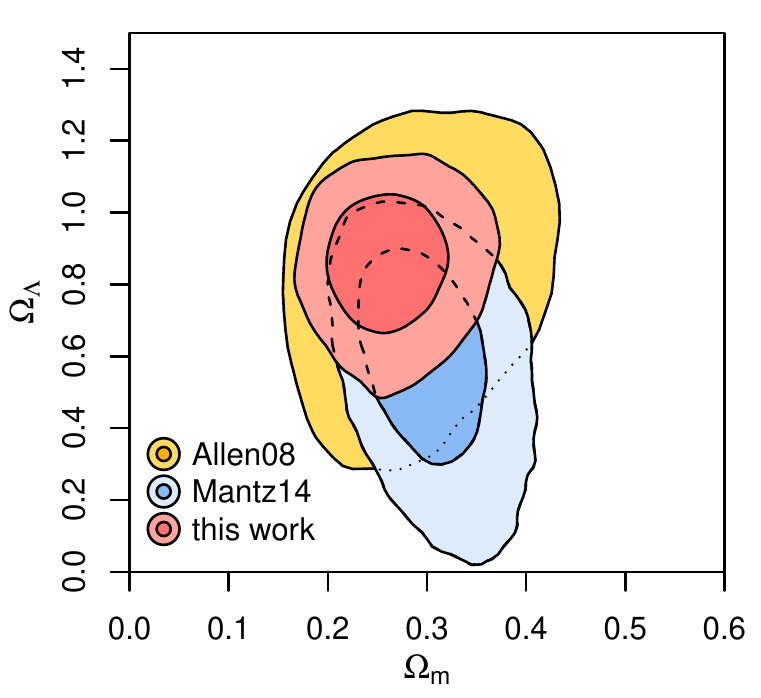}
  \hspace{10mm}
  \includegraphics{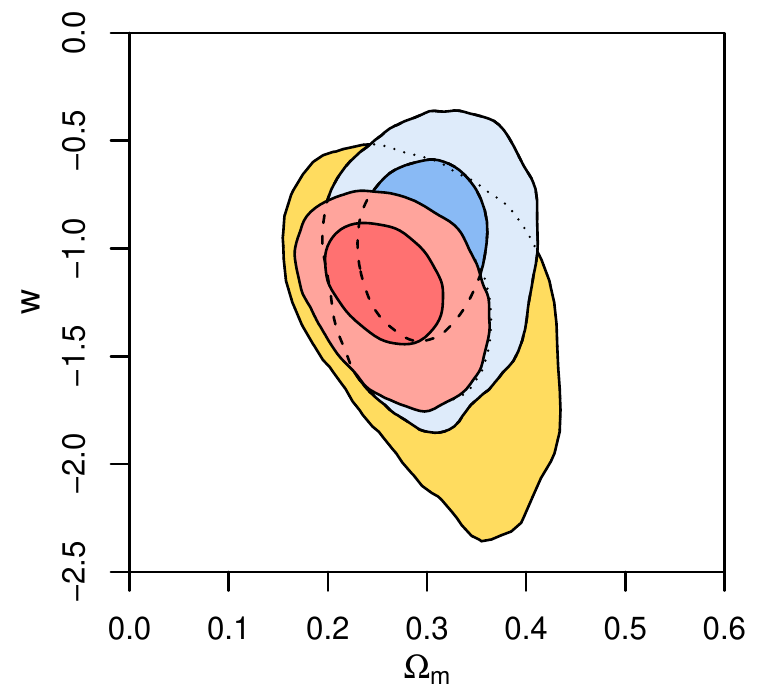}
  \caption{
    Constraints on parameters of the \LCDM{} (left) and flat, constant-$w$ (right) models from 3 generations of the \fgas{} analysis presented in this paper: \citet{Allen0706.0033}, \mamrakls{}, and this work.
    Compared with \mamrakls{}, our new constraints on $\Omegal$ are 41 per cent tighter, while constraints on $w$ are 29 per cent tighter.
  }
  \label{fig:improvement}
\end{figure*}

Our results are consistent with flat \LCDM{}, and in good agreement individually with those from CMB, SN, BAO and $3\times2$pt data, as shown in Figure~\ref{fig:lcdm}, though we note that the CMB and BAO constraints shown appear to be in tension with one another for the flat, constant-$w$ model (this is not true of the older BAO data set that we employ in the combined-probe analysis below). 
For the constant-$w$ model in particular, our constraint of $w=-1.13^{+0.17}_{-0.20}$ compares well with those from other low-redshift probes: $w=-1.09\pm0.22$ (SN), $w=-0.69\pm0.15$ (BAO), and $w=-0.98^{+0.32}_{-0.20}$ ($3\times2$pt).
The \fgas{} results are also comparable to independent constraints from the number counts of massive clusters ($w=-1.01\pm0.20$; \citealt{Mantz0909.3098}; see also \citealt{Vikhlinin0812.2720, Mantz1407.4516, Bocquet1812.01679}).
Moreover, apart from a very weak dependence on $h$ (Section~\ref{sec:model}), the \fgas{} data provide these constraints without sensitivity to (or requiring assumptions about) additional parameters such as the number or mass of neutrinos, or the shape or amplitude of the matter power spectrum, which do impact some of the probes discussed above.

\subsection{Combination with independent probes}

\begin{figure*}
  \centering
  \includegraphics{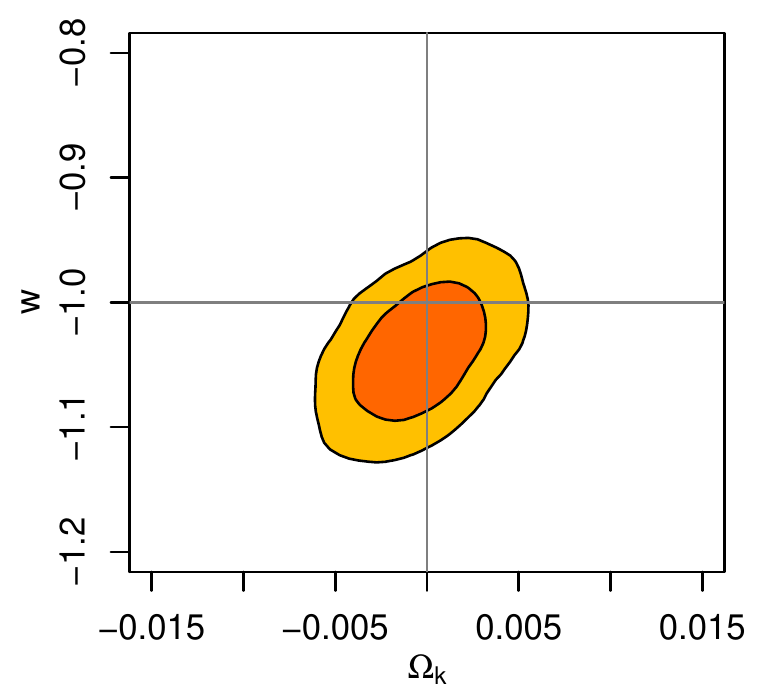}
  \hspace{10mm}
  \includegraphics{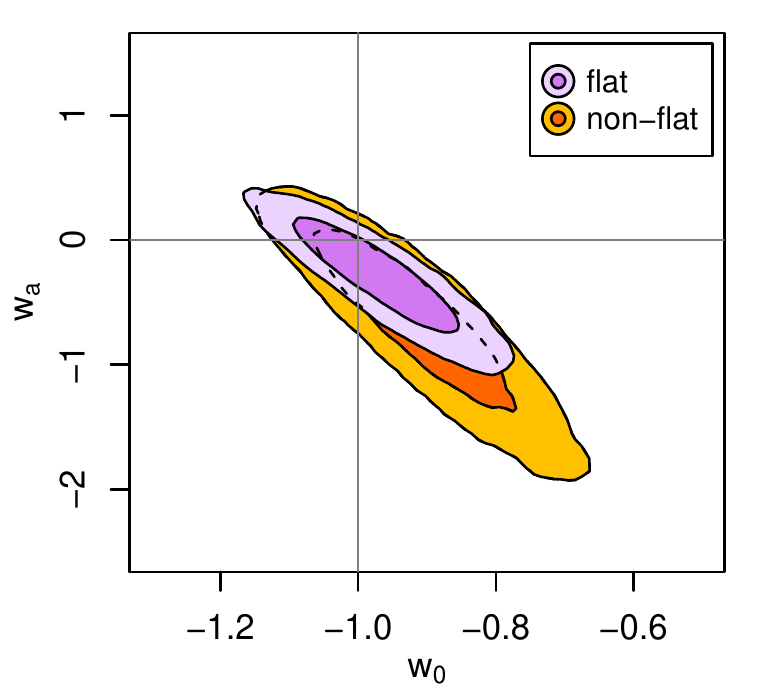}
  \caption{
    Constraints on parameters of the non-flat, constant-$w$ (left) and evolving-$w$ (right) models from the combination of \fgas{} (this work), \Planck{} CMB \citep{Planck1907.12875, Planck1807.06210}, Pantheon SN \citep{Scolnic1710.00845}, and BAO \citep{Beutler1106.3366, Ross1409.3242, Alam1607.03155} data.
    Gray lines correspond to the values each parameter takes in the flat \LCDM{} model.
  }
  \label{fig:okw}
  \label{fig:evolw}
\end{figure*}

Combining the \fgas{} data with CMB, SN and BAO information, we obtain the constraints listed in Table~\ref{tab:comb_constraints}.
Note that the priors on $h$ and $\Omegab h^2$ employed in the last section are not used here, as the above combination of data is sufficient to constrain those parameters tightly in all of the models considered here.
We find $\Omegak=(0.9\pm2.0)\E{-3}$ for the non-flat \LCDM{} model, and $w=-1.04\pm0.03$ for the flat, constant-$w$ model, with $\Omegam$ tightly constrained around values of 0.30--0.31 in both cases.
Generalizing to a model with both curvature and the dark energy equation of state free, the combination yields $\Omegak=(-0.5\pm2.2)\E{-3}$ and $w=-1.04\pm0.04$ (left panel of Figure~\ref{fig:okw}).

\begin{table*}
  \begin{center}
    \caption{
      Marginalized best-fitting values and 68.3 per cent probability credible intervals on cosmological parameters of constant- and evolving-$w$ models from the combination of \fgas{} (this work), \Planck{} CMB \citep{Planck1907.12875, Planck1807.06210}, Pantheon SN \citep{Scolnic1710.00845}, and BAO \citep{Beutler1106.3366, Ross1409.3242, Alam1607.03155} data.
    }
    \label{tab:comb_constraints}
    \vspace{1ex}
    \begin{tabular}{ccccccc}
      \hline
      Model & $h$ & \Omegam{} & \Omegade{} & \phmin$10^3\Omega_k$ & $w_0$ & \phmin$w_a$ \\
      \hline\vspace{-1.5ex}\\
      \input{tables/comb_constraints}
      \hline
    \end{tabular}
  \end{center}
\end{table*}

We also consider models with an evolving equation of state, of the form
\begin{equation} \label{eq:wz}
  w(z) = w_0 + w_a(1-a) = w_0 + w_a\left(\frac{z}{1+z}\right),
\end{equation}
where $a=(1+z)^{-1}$ is the scale factor and $w_a$ parametrizes the change in $w(z)$ between the present day ($a=1$) and the early Universe ($a\rightarrow0$).
The right panel of Figure~\ref{fig:evolw} shows constraints on $w_0$ and $w_a$, with and without free curvature, from the combined data set.
In the former case, we find $\Omegak=(-3.3\pm3.0)\E{-3}$, $w_0=-0.92\pm0.10$ and $w_a=-0.60\pm0.47$, consistent with the flat \LCDM{} model.

\subsection{Constraints on cluster parameters} \label{sec:misc}

Beyond the cosmological parameters, there are 4 parameters of the cluster model that the data constrain.
These constraints are essentially independent of cosmological assumptions within the range of models explored above;
the specific values quoted below are from a fit of the flat, constant-$w$ model, using the $\fgas$ data plus priors on $h$ and $\Omegab h^2$.

The power-law slope of $\fgas$ in the 0.8--1.2\,$r_{2500}$ shell with mass is found to be $\alpha=0.025\pm0.033$, consistent with zero.
While the gas mass fraction is expected to be an increasing function of mass going from the group to the cluster regime, our results verify that this trend becomes consistent with a constant for sufficiently massive clusters at these radii \citep{Eke9708070, Kay0407058, Crain0610602, Nagai0609247, Young1007.0887, Battaglia1209.4082, Planelles1209.5058, Le-Brun1312.5462, Le-Brun1606.04545, Barnes1607.04569, Henden1911.12367, Singh1911.05751}.
We note that there is visually no indication of steepening towards lower masses in Figure~\ref{fig:fgas_zM}.

For the log-normal intrinsic scatter in $\fgas$, we find $\sigma_f=0.043^{+0.020}_{-0.032}$, with a 95.4 per cent probability upper limit of $0.089$.
This result is consistent with, though smaller than, the previous constraint of $0.074\pm0.023$ from \mamrakls{} (see also \citealt{Mahdavi1210.3689, Herbonnet1912.04414}).
Remarkably, this best-fitting scatter of just 4.3 per cent in $\fgas$ corresponds to a precision of 2.9 per cent in the distance estimate associated with a given cluster, compared with $\sim4.6$ per cent intrinsic scatter in distance estimates from type Ia supernovae \citep{Scolnic1710.00845}.

We constrain the weak lensing to X-ray mass ratio to be $K_0=0.93\pm0.11$, a small change from \amalmhkbers.
Strictly speaking, this parameter describes the ratio at $z=0$, but the posterior correlation between $K_0$ and the evolution parameter, $K_1$, is small enough that the constraints on the ratio at other relevant redshifts ($z<1.2$) are identical at this precision.
$K_1$ itself is unconstrained by the data.
Note that, unlike the other parameters discussed here, $K_0$ is degenerate with cosmological parameters, primarily $\Omegam$.
We therefore obtain tighter and slightly offset constraints from the combination of $\fgas$, CMB, BAO and SN data, which prefer a larger value of $\Omegam$ than \fgas{} alone: $K_0=0.99\pm0.06$.
Finally, the intrinsic scatter in the lensing to X-ray mass ratio is found to be $\sigma_K = 0.14_{-0.07}^{+0.09}$, consistent with expectations for the impact of halo triaxiliaty on the inferred 3D lensing masses \citep{Becker1011.1681}.

\section{Conclusion} \label{sec:conclusion}

We have derived constraints on cosmological models from measurements of the gas mass fraction at intermediate radii in a sample of the most dynamically relaxed, massive galaxy clusters.
Compared to previous work, our analysis incorporates additional \Chandra{} data, as well as an expanded cluster sample.
Notably, the expanded data set include a precise determination of \fgas{} in the Perseus Cluster, at $z=0.018$, as well as new measurements at $z=0.97$ and $z=1.16$.
The resulting increase in leverage on the apparent evolution of \fgas{} with redshift has a disproportionate impact on dark energy constraints; compared with \mamrakls{}, we find that the constraint on $\Omegal$ for the non-flat \LCDM{} model shrinks by 41 per cent to $\Omegal=0.865\pm0.119$, while the constraint on $w$ in the flat, constant-$w$ model improves by 29 per cent to $w=-1.13_{-0.20}^{+0.17}$.
Despite the modest size of the data set overall, comprising observations of just 44 sources, dark energy constraints from \fgas{} data remain competitive with the best constraints from other cosmological probes.
Combining the \fgas{} analysis with CMB, SN and BAO data, we explore non-flat and evolving-$w$ models, continuing to find consistency with the simple flat \LCDM{} concordance model.
The lowest-redshift \fgas{} data, combined only with a measurement of $\Omegab/\Omegam$ from the CMB, constrain the Hubble constant to be $h=0.722\pm0.067$.

Our analysis internally constrains a possible trend with mass and the intrinsic scatter of \fgas.
We obtain tight constraints on the power-law slope of \fgas{} (in a shell including radii of 0.8--1.2\,$r_{2500}$) with mass, $\alpha=0.025\pm0.033$, consistent with no dependence in this intermediate radial range, given the cluster selection.
The log-normal intrinsic scatter is found to be $\sigma_f=0.043^{+0.020}_{-0.032}$, somewhat smaller than in previous work and equivalent to just 3 per cent scatter in the distance estimated to a single cluster.

Even with the additional data introduced in this work, constraints on dark energy from the \fgas{} method remain statistically limited.
With the precise anchor at low redshifts provided by Perseus and other bright clusters, long-term improvement will be driven by additions to the cluster sample at $z\gtsim0.5$.
The growing number of relaxed clusters in our data set at $0.6<z<1.2$ that were discovered via the SZ effect \citep{Bleem1409.0850, Bleem1910.04121, Planck1502.01598}, now 6 in total, demonstrates that this is a viable approach to finding such systems, even if a smaller fraction of relaxed clusters is found in SZ surveys at these redshifts compared with X-ray surveys of the low-redshift Universe.
In the long term, significant expansions of the sample have the potential to dramatically improve constraints on dark energy, as discussed by \mamrakls{}.
In the nearer term, dark-energy-independent constraints on $\Omegam$ can be straightforwardly tightened by expanding and improving the weak lensing data for relaxed clusters.

\section*{Acknowledgements}

We thank the referee for their careful reading of the manuscript.
The scientific results reported in this article are based on observations made by the Chandra X-ray Observatory, and data obtained from the Chandra Data Archive.
Support for this work was provided by the National Aeronautics and Space Administration through Chandra Awards Number GO5-16122X, GO5-16148X, GO6-17112B, and GO0-21124B issued by the Chandra X-ray Center, which is operated by the Smithsonian Astrophysical Observatory for and on behalf of the National Aeronautics Space Administration under contract NAS8-03060.
This research has made use of software provided by the Chandra X-ray Center (CXC) in the application packages CIAO.
We acknowledge support from the U.S. Department of Energy under contract numbers DE-AC02-76SF00515 and DE-SC0018053.
PLK acknowledges support from NSF AST-1908823.
NW is supported by the GACR grant 21-13491X.

This work was performed in the context of the South Pole Telescope scientific program. SPT is supported by the National Science Foundation through grants PLR-1248097 and OPP-1852617.
Partial support is also provided by the NSF Physics Frontier Center grant PHY-1125897 to the Kavli Institute of Cosmological Physics at the University of Chicago, the Kavli Foundation and the Gordon and Betty Moore Foundation grant GBMF 947 to the University of Chicago.
The SPT is also supported by the U.S.\ Department of Energy.
Work at Argonne National Lab was supported by the U.S.\ Department of Energy, Office of Science, Office of High Energy Physics, under contract DE-AC02- 06CH11357.
Work at Fermi National Accelerator Laboratory, a DOE-OS, HEP User Facility managed by the Fermi Research Alliance, LLC, was supported under Contract No. DE-AC02-07CH11359.

%%%%%%%%%%%%%%%%%%%%%%%%%%%%%%%%%%%%%%%%%%%%%%%%%%
\section*{Data Availability}

\Chandra{} X-ray data are available from the  \Chandra{} Data Archive (CDA) at \url{https://cxc.harvard.edu/cda/}.
Appendix~\ref{sec:data_table} provides the specific observation IDs employed in this work.
Our weak lensing analysis uses data from the Subaru and Canada-France-Hawaii telescopes (see \citealt{von-der-Linden1208.0597} and \amalmhkbers{} for details), which can respectively be obtained from the SMOKA Science Archive (\url{https://smoka.nao.ac.jp/index.jsp}) and the Canadian Astronomy Data Centre (\url{https://www.cadc-ccda.hia-iha.nrc-cnrc.gc.ca/en/cfht/index.html}).
The code and data tables required to perform the \fgas{} cosmological analysis can be obtained from \url{https://github.com/abmantz/fgas-cosmo}.
All data shown in figures and tables can be obtained in digital form from \url{https://github.com/abmantz/fgas-2021-paper}.

%The inclusion of a Data Availability Statement is a requirement for articles published in MNRAS. Data Availability Statements provide a standardised format for readers to understand the availability of data underlying the research results described in the article. The statement may refer to original data generated in the course of the study or to third-party data analysed in the article. The statement should describe and provide means of access, where possible, by linking to the data or providing the required accession numbers for the relevant databases or DOIs.



\def \aap {A\&A}
\def \aapr {A\&AR}
\def \apj {ApJ}
\def \apjl {ApJ}
\def \apjs {ApJS}
\def \araa {ARA\&A}
\def \asl {Adv.\ Sci.\ Lett.}
\def \mnras {MNRAS}
\def \nat {Nat}
\def \science {Sci}

%\begin{thebibliography}{99}
%\end{thebibliography}

%%%%%%%%%%%%%%%%%%%%%%%%%%%%%%%%%%%%%%%%%%%%%%%%%%

%%%%%%%%%%%%%%%%% APPENDICES %%%%%%%%%%%%%%%%%%%%%

\appendix

\section{Chandra Data} \label{sec:data_table}

Table~\ref{tab:obs} lists details of the \Chandra{} data used in this work.

\begin{table*}
  \centering
  \caption{
    Chandra data used in this work:
    [1] cluster name (ordered by increasing redshift);
    [2] observation ID;
    [3] date of observation;
    [4] detector (ACIS-I or ACIS-S);
    [5] clean exposure time in ks.
  }
  \begin{tabular}{lrccr@{\hspace{20mm}}lrccr}
    \hline
    Cluster & ObsId & Date & Det & Exp & Cluster & ObsId & Date & Det & Exp \\
    \hline\vspace{-3ex}\\
    \input{tables/observation_info}
    \hline
  \end{tabular}
  \label{tab:obs}
\end{table*}

\begin{table*}
  \centering
  \contcaption{}
    \begin{tabular}{lrccr@{\hspace{20mm}}lrccr}
    \hline
      Cluster & ObsId & Date & Det & Exp & Cluster & ObsId & Date & Det & Exp \\
    \hline\vspace{-3ex}\\
    \input{tables/observation_info2}
    \hline
  \end{tabular}
\end{table*}

\section{Model parameters and priors}

Tables~\ref{tab:fgas_params}--\ref{tab:theta_params} list the astrophysical and cosmological model parameters of our analysis, and the prior distributions adopted for each.

\begin{table*}
  \centering
  \caption{
    Parameters and priors specific to the cluster \fgas{} model (see Section~\ref{sec:model}).
    $\mathcal{N}(\mu,\sigma)$ represents the normal distribution with mean $\mu$ and variance $\sigma^2$, and $\mathcal{U}(x_1,x_2)$ the uniform distribution with endpoints $x_1$ and $x_2$.
    Priors marked with a $\star$ are informative in the sense that the posterior distribution for the corresponding parameter essentially reproduces the prior (see Section~\ref{sec:model} and \mamrakls{} for discussion).
    Note that in Section~\ref{sec:om} we use a much wider prior on $\Upsilon_0$ than that listed here.
  }
  \vspace{1ex}
  \begin{tabular}{lll}
    \hline
    Symbol & Meaning & Prior \\
    \hline\vspace{-1.5ex}\\
    \input{tables/fgas_parameters}
    \hline
  \end{tabular}
  \label{tab:fgas_params}
\end{table*}

\begin{table*}
  \centering
  \caption{
    Parameters and priors of the cosmological model used when analyzing the $\fgas$, SN or BAO data alone, presented as in Table~\ref{tab:fgas_params}.
    $\Omegab h^2$ is derived from the other parameters listed, rather than being a free parameter itself.
    Note that the normal priors on $H_0$ and $\Omegab h^2$ (indicated by the $\dagger$ symbol) are not used for results in Sections~\ref{sec:lowz} and \ref{sec:om}, nor in any joint analysis of \fgas{} and CMB data.
  }
  \vspace{1ex}
  \begin{tabular}{lll}
    \hline
    Symbol & Meaning & Prior \\
    \hline\vspace{-1.5ex}\\
    \input{tables/astro_parameters}
    \hline
  \end{tabular}
  \label{tab:astro_params}
\end{table*}

\begin{table*}
  \centering
  \caption{
    Parameters and priors of the cosmological model used when analyzing CMB data, alone or in combination with other probes, presented as in Table~\ref{tab:fgas_params}.
    Adopted values of the neutrino mass and effective number of relativistic species are shown for reference, but were not varied.
    $H_0$ is derived from the other parameters listed, rather than being a free parameter itself.
  }
  \vspace{1ex}
  \begin{tabular}{lll}
    \hline
    Symbol & Meaning & Prior \\
    \hline\vspace{-1.5ex}\\
    \input{tables/theta_parameters}
    \hline
  \end{tabular}
  \label{tab:theta_params}
\end{table*}

%%%%%%%%%%%%%%%%%%%%%%%%%%%%%%%%%%%%%%%%%%%%%%%%%%

% Don't change these lines
%\bsp	% typesetting comment
\label{lastpage}
\end{document}

%% file: tables/fgas_mass_table.tex
Perseus                   &  0.018  &  1.638  &  0.014  &  1.30  &  0.04  &  0.129  &  0.004  &  2.66   &  0.04  \\
Abell~2029                &  0.078  &  2.243  &  0.021  &  1.69  &  0.06  &  0.134  &  0.004  &  4.06   &  0.09  \\
Abell~478                 &  0.088  &  2.273  &  0.043  &  1.94  &  0.17  &  0.117  &  0.009  &  4.20   &  0.24  \\
RX~J1524.2$-$3154         &  0.103  &  2.740  &  0.030  &  2.07  &  0.08  &  0.132  &  0.004  &  4.58   &  0.13  \\
PKS~0745$-$191            &  0.103  &  0.827  &  0.022  &  0.67  &  0.06  &  0.123  &  0.009  &  1.76   &  0.09  \\
Abell~2204$^\ast$         &  0.152  &  2.719  &  0.055  &  1.95  &  0.15  &  0.139  &  0.008  &  5.09   &  0.24  \\
RX~J0439.0+0520           &  0.208  &  0.853  &  0.037  &  0.74  &  0.13  &  0.115  &  0.017  &  1.98   &  0.19  \\
Zwicky~2701               &  0.214  &  0.934  &  0.019  &  0.83  &  0.06  &  0.112  &  0.006  &  1.88   &  0.08  \\
RX~J1504.1$-$0248         &  0.215  &  2.571  &  0.027  &  1.93  &  0.09  &  0.133  &  0.005  &  4.92   &  0.16  \\
Zwicky~2089$^\ast$        &  0.235  &  1.815  &  0.060  &  1.18  &  0.16  &  0.154  &  0.019  &  2.85   &  0.20  \\
RX~J2129.6+0005           &  0.235  &  0.829  &  0.026  &  0.72  &  0.08  &  0.115  &  0.011  &  1.64   &  0.12  \\
RX~J1459.4$-$1811         &  0.236  &  1.833  &  0.076  &  1.21  &  0.13  &  0.151  &  0.014  &  2.73   &  0.23  \\
Abell~1835$^\ast$         &  0.252  &  3.088  &  0.040  &  2.38  &  0.13  &  0.130  &  0.006  &  4.99   &  0.15  \\
Abell~3444                &  0.253  &  2.165  &  0.071  &  1.53  &  0.17  &  0.141  &  0.012  &  3.11   &  0.20  \\
MS~2137.3$-$2353$^\ast$   &  0.313  &  1.093  &  0.031  &  0.79  &  0.07  &  0.138  &  0.011  &  2.09   &  0.12  \\
MACS~J0242.5$-$21         &  0.314  &  1.421  &  0.123  &  1.09  &  0.30  &  0.130  &  0.029  &  2.90   &  0.53  \\
MACS~J1427.6$-$25         &  0.318  &  0.814  &  0.047  &  0.65  &  0.13  &  0.126  &  0.022  &  1.80   &  0.23  \\
MACS~J2229.7$-$27         &  0.324  &  1.170  &  0.050  &  0.81  &  0.13  &  0.144  &  0.021  &  1.98   &  0.20  \\
MACS~J0947.2+7623         &  0.345  &  2.042  &  0.075  &  1.90  &  0.23  &  0.107  &  0.010  &  4.43   &  0.37  \\
MACS~J1931.8$-$26         &  0.352  &  1.943  &  0.043  &  1.48  &  0.13  &  0.131  &  0.010  &  3.45   &  0.18  \\
MACS~J1115.8+0129$^\ast$  &  0.355  &  2.170  &  0.055  &  1.39  &  0.15  &  0.156  &  0.014  &  3.38   &  0.24  \\
MACS~J1532.8+3021$^\ast$  &  0.363  &  0.921  &  0.039  &  0.58  &  0.09  &  0.159  &  0.021  &  1.51   &  0.17  \\
MACS~J0150.3$-$10         &  0.363  &  1.912  &  0.053  &  1.69  &  0.15  &  0.113  &  0.007  &  3.62   &  0.22  \\
RCS~J1447+0828            &  0.376  &  2.401  &  0.070  &  1.83  &  0.16  &  0.131  &  0.008  &  3.99   &  0.28  \\
MACS~J0011.7$-$15         &  0.378  &  1.617  &  0.070  &  1.20  &  0.21  &  0.135  &  0.021  &  3.02   &  0.30  \\
MACS~J1720.2+3536$^\ast$  &  0.391  &  1.681  &  0.053  &  1.24  &  0.18  &  0.136  &  0.016  &  3.09   &  0.34  \\
MACS~J0429.6$-$02$^\ast$  &  0.399  &  1.579  &  0.126  &  1.52  &  0.43  &  0.104  &  0.023  &  3.21   &  0.61  \\
MACS~J0159.8$-$08         &  0.404  &  2.584  &  0.107  &  2.10  &  0.34  &  0.123  &  0.017  &  4.91   &  0.47  \\
MACS~J2046.0$-$34         &  0.423  &  1.046  &  0.052  &  0.67  &  0.13  &  0.156  &  0.026  &  1.71   &  0.20  \\
IRAS~09104+4109           &  0.442  &  1.243  &  0.066  &  1.16  &  0.21  &  0.107  &  0.015  &  2.82   &  0.33  \\
MACS~J1359.1$-$19         &  0.447  &  0.925  &  0.066  &  0.96  &  0.23  &  0.096  &  0.019  &  2.22   &  0.36  \\
RX~J1347.5$-$1145$^\ast$  &  0.451  &  4.883  &  0.106  &  4.23  &  0.37  &  0.116  &  0.008  &  10.93  &  0.68  \\
3C~295                    &  0.460  &  1.024  &  0.063  &  0.91  &  0.21  &  0.112  &  0.021  &  2.12   &  0.31  \\
MACS~J1621.3+3810$^\ast$  &  0.461  &  1.390  &  0.048  &  1.10  &  0.16  &  0.126  &  0.016  &  2.76   &  0.23  \\
MACS~J1427.2+4407$^\ast$  &  0.487  &  1.546  &  0.089  &  0.97  &  0.17  &  0.159  &  0.022  &  2.40   &  0.30  \\
MACS~J1423.8+2404$^\ast$  &  0.539  &  1.552  &  0.054  &  1.12  &  0.13  &  0.138  &  0.012  &  2.83   &  0.21  \\
SPT~J2331$-$5051          &  0.576  &  1.120  &  0.039  &  0.92  &  0.07  &  0.121  &  0.006  &  2.07   &  0.17  \\
SPT~J2344$-$4242          &  0.596  &  3.044  &  0.076  &  2.48  &  0.15  &  0.122  &  0.004  &  6.07   &  0.33  \\
SPT~J0000$-$5748          &  0.702  &  0.944  &  0.036  &  0.93  &  0.14  &  0.102  &  0.012  &  2.05   &  0.23  \\
SPT~J2043$-$5035          &  0.723  &  1.210  &  0.037  &  1.11  &  0.11  &  0.109  &  0.008  &  2.19   &  0.18  \\
SPT~J0615$-$5746          &  0.972  &  2.969  &  0.075  &  2.54  &  0.22  &  0.117  &  0.007  &  4.70   &  0.36  \\
CL~J1415.2+3612           &  1.028  &  0.846  &  0.024  &  0.98  &  0.18  &  0.087  &  0.015  &  1.92   &  0.26  \\
3C~186                    &  1.063  &  0.992  &  0.067  &  0.97  &  0.22  &  0.102  &  0.018  &  1.91   &  0.30  \\
SPT~J2215$-$3537          &  1.160  &  1.569  &  0.085  &  1.07  &  0.20  &  0.147  &  0.021  &  2.40   &  0.35  \\

%% file: tables/lowz_constraints.tex
--- & --- & --- & $h^{3/2}\Omegab/\Omegam$ & $0.096$ & $0.013$\vspace{1ex}\\
--- & --- & P18 & $h$ & $0.722$ & $0.067$\vspace{1ex}\\
F01 & --- & --- & $\Omegab/\Omegam$ & $0.156$ & $0.034$\vspace{1ex}\\
P18 & --- & --- & $\Omegab/\Omegam$ & $0.173$ & $0.024$\vspace{1ex}\\
R19 & --- & --- & $\Omegab/\Omegam$ & $0.150$ & $0.021$\vspace{1ex}\\
--- & H20 & --- & $\Omegam h^{1/2}$ & $0.221$ & $0.031$\vspace{1ex}\\
F01 & H20 & --- & $\Omegam$ & $0.260$ & $0.040$\vspace{1ex}\\
P18 & H20 & --- & $\Omegam$ & $0.270$ & $0.038$\vspace{1ex}\\
R19 & H20 & --- & $\Omegam$ & $0.257$ & $0.037$\vspace{1ex}\\

%% file: tables/fgas_constraints.tex
\LCDM{} & $0.257\pm0.039$ & $0.865\pm0.119$ & $-0.128\pm0.128$ & $-1$\vspace{1ex}\\
constant-$w$ & $0.256\pm0.037$ & $0.744\pm0.037$ & \phmin0 & $-1.13^{+0.17}_{-0.20}$\vspace{1ex}\\

%% file: tables/comb_constraints.tex
\LCDM{} & $0.680\pm0.007$ & $0.308\pm0.006$ & $0.691\pm0.005$ & \phmin$0.9\pm2.0$ & $-1$ &  \phmin0\vspace{1ex}\\
constant-$w$ & $0.686\pm0.008$ & $0.303\pm0.007$ & $0.697\pm0.007$ & \phmin0 & $-1.04\pm0.03$ &  \phmin0\vspace{1ex}\\
& $0.686\pm0.009$ & $0.303\pm0.008$ & $0.698\pm0.008$ & $-0.5\pm2.2$ & $-1.04\pm0.04$ &  \phmin0\vspace{1ex}\\
evolving-$w$ & $0.686\pm0.008$ & $0.303\pm0.007$ & $0.697\pm0.007$ & \phmin0 & $-0.97\pm0.08$ &  $-0.27\pm0.30$\vspace{1ex}\\
& $0.683\pm0.009$ & $0.304\pm0.008$ & $0.699\pm0.008$ & $-3.3\pm3.0$ & $-0.92\pm0.10$ &  $-0.60\pm0.47$\vspace{1ex}\\

%% file: tables/observation_info.tex
Perseus            &  3237   &  2003-03-15  &  S    &  33.8   &    RX~J1504.1$-$0248  &  17197  &  2015-06-01  &  I    &  24.2   \\
~                  &  4953   &  2004-10-18  &  S    &  30.1   &    ~                  &  17669  &  2015-06-17  &  I    &  24.9   \\
~                  &  6145   &  2004-10-19  &  S    &  85.0   &    ~                  &  17670  &  2015-06-10  &  I    &  42.2   \\
~                  &  6146   &  2004-10-20  &  S    &  28.7   &    Zwicky~2089        &  7897   &  2006-12-23  &  I    &  8.4    \\
~                  &  11713  &  2009-11-29  &  I    &  108.4  &    ~                  &  10463  &  2009-02-24  &  S    &  38.6   \\
~                  &  11714  &  2009-12-07  &  I    &  89.7   &    RX~J2129.6+0005    &  552    &  2000-10-21  &  I    &  9.4    \\
~                  &  11715  &  2009-12-02  &  I    &  70.0   &    ~                  &  9370   &  2009-04-03  &  I    &  27.3   \\
~                  &  11716  &  2009-10-10  &  I    &  38.6   &    RX~J1459.4$-$1811  &  9428   &  2008-06-16  &  S    &  39.6   \\
~                  &  12025  &  2009-11-25  &  I    &  17.9   &    Abell~1835         &  496    &  2000-04-29  &  S    &  10.7   \\
~                  &  12033  &  2009-11-27  &  I    &  18.4   &    ~                  &  6880   &  2006-08-25  &  I    &  106.7  \\
~                  &  12036  &  2009-12-02  &  I    &  46.8   &    ~                  &  6881   &  2005-12-07  &  I    &  29.9   \\
~                  &  12037  &  2009-12-05  &  I    &  83.1   &    ~                  &  7370   &  2006-07-24  &  I    &  36.3   \\
~                  &  13989  &  2011-11-07  &  I    &  34.3   &    Abell~3444         &  9400   &  2008-02-11  &  S    &  35.7   \\
~                  &  13990  &  2011-11-11  &  I    &  34.0   &    MS~2137.3$-$2353   &  928    &  1999-11-18  &  S    &  23.3   \\
~                  &  13991  &  2011-11-05  &  I    &  32.7   &    ~                  &  5250   &  2003-11-18  &  S    &  27.6   \\
~                  &  13992  &  2011-11-05  &  I    &  32.9   &    MACS~J0242.5$-$21  &  3266   &  2002-02-07  &  I    &  7.7    \\
~                  &  17260  &  2015-12-01  &  I    &  4.7    &    MACS~J1427.6$-$25  &  3279   &  2002-06-29  &  I    &  14.4   \\
~                  &  17261  &  2015-12-01  &  I    &  4.2    &    ~                  &  9373   &  2008-06-11  &  I    &  26.9   \\
~                  &  17262  &  2015-12-07  &  I    &  4.2    &    MACS~J2229.7$-$27  &  3286   &  2002-11-13  &  I    &  11.5   \\
~                  &  17263  &  2015-12-04  &  I    &  4.3    &    ~                  &  9374   &  2007-12-09  &  I    &  14.3   \\
~                  &  17274  &  2015-12-09  &  I    &  5.0    &    MACS~J0947.2+7623  &  2202   &  2000-10-20  &  I    &  10.7   \\
~                  &  17275  &  2015-12-09  &  I    &  4.7    &    ~                  &  7902   &  2007-07-09  &  S    &  38.3   \\
~                  &  17276  &  2015-12-09  &  I    &  4.8    &    MACS~J1931.8$-$26  &  3282   &  2002-10-20  &  I    &  11.5   \\
~                  &  17277  &  2015-12-10  &  I    &  4.6    &    ~                  &  9382   &  2008-08-21  &  I    &  92.5   \\
~                  &  17278  &  2015-12-10  &  I    &  4.4    &    MACS~J1115.8+0129  &  3275   &  2003-01-23  &  I    &  9.5    \\
~                  &  17279  &  2015-11-30  &  I    &  4.6    &    ~                  &  9375   &  2008-02-03  &  I    &  34.8   \\
~                  &  17280  &  2015-12-11  &  I    &  4.5    &    MACS~J1532.8+3021  &  1649   &  2001-08-26  &  S    &  9.2    \\
~                  &  17283  &  2015-10-06  &  I    &  5.0    &    ~                  &  1665   &  2001-09-06  &  I    &  8.4    \\
~                  &  17286  &  2015-10-06  &  I    &  4.5    &    ~                  &  14009  &  2011-11-16  &  S    &  84.8   \\
Abell~2029         &  891    &  2000-04-12  &  S    &  19.8   &    MACS~J0150.3$-$10  &  11711  &  2009-09-14  &  I    &  26.1   \\
~                  &  4977   &  2004-01-08  &  S    &  72.8   &    RCS~J1447+0828     &  10481  &  2008-12-14  &  S    &  9.2    \\
~                  &  6101   &  2004-12-17  &  I    &  8.7    &    ~                  &  17233  &  2016-04-05  &  I    &  37.9   \\
~                  &  10434  &  2009-04-01  &  I    &  4.9    &    ~                  &  18825  &  2016-04-06  &  I    &  21.9   \\
~                  &  10435  &  2009-04-01  &  I    &  4.1    &    MACS~J0011.7$-$15  &  3261   &  2002-11-20  &  I    &  17.5   \\
~                  &  10436  &  2009-04-01  &  I    &  4.1    &    ~                  &  6105   &  2005-06-28  &  I    &  31.7   \\
~                  &  10437  &  2009-04-01  &  I    &  4.5    &    MACS~J1720.2+3536  &  3280   &  2002-11-03  &  I    &  18.3   \\
Abell~478          &  1669   &  2001-01-27  &  S    &  36.2   &    ~                  &  6107   &  2005-11-22  &  I    &  27.2   \\
~                  &  6102   &  2004-09-13  &  I    &  5.9    &    ~                  &  7718   &  2007-09-28  &  I    &  6.2    \\
~                  &  6928   &  2005-12-02  &  I    &  5.7    &    MACS~J0429.6$-$02  &  3271   &  2002-02-07  &  I    &  19.3   \\
~                  &  6929   &  2005-12-02  &  I    &  1.9    &    MACS~J0159.8$-$08  &  3265   &  2002-10-02  &  I    &  14.6   \\
~                  &  7217   &  2005-11-15  &  I    &  17.0   &    ~                  &  6106   &  2004-12-04  &  I    &  31.0   \\
~                  &  7218   &  2005-11-17  &  I    &  6.6    &    ~                  &  9376   &  2008-10-03  &  I    &  17.0   \\
~                  &  7222   &  2005-11-19  &  I    &  5.4    &    MACS~J2046.0$-$34  &  5816   &  2005-06-28  &  I    &  8.2    \\
~                  &  7231   &  2006-07-29  &  I    &  15.4   &    ~                  &  9377   &  2008-06-27  &  I    &  34.6   \\
~                  &  7232   &  2005-12-04  &  I    &  13.0   &    IRAS~09104+4109    &  10445  &  2009-01-06  &  I    &  69.0   \\
~                  &  7233   &  2005-12-03  &  I    &  7.7    &    MACS~J1359.1$-$19  &  5811   &  2005-03-17  &  I    &  8.9    \\
~                  &  7234   &  2005-12-01  &  I    &  7.8    &    ~                  &  9378   &  2008-03-21  &  I    &  45.3   \\
~                  &  7235   &  2005-11-29  &  I    &  6.8    &    RX~J1347.5$-$1145  &  506    &  2000-03-05  &  S    &  8.2    \\
RX~J1524.2$-$3154  &  9401   &  2008-01-07  &  S    &  40.9   &    ~                  &  507    &  2000-04-29  &  S    &  10.0   \\
PKS~0745$-$191     &  2427   &  2001-06-16  &  S    &  17.9   &    ~                  &  3592   &  2003-09-03  &  I    &  48.1   \\
~                  &  6103   &  2004-09-24  &  I    &  9.2    &    ~                  &  13516  &  2012-12-11  &  I    &  34.4   \\
~                  &  7694   &  2007-01-25  &  I    &  4.7    &    ~                  &  13999  &  2012-05-14  &  I    &  50.8   \\
~                  &  12881  &  2011-01-27  &  S    &  117.0  &    ~                  &  14407  &  2012-03-16  &  I    &  55.0   \\
Abell~2204         &  499    &  2000-07-29  &  S    &  9.0    &    3C~295             &  578    &  1999-08-30  &  S    &  15.4   \\
~                  &  6104   &  2004-09-20  &  I    &  8.6    &    ~                  &  2254   &  2001-05-18  &  I    &  75.5   \\
~                  &  7940   &  2007-06-06  &  I    &  72.5   &    MACS~J1621.3+3810  &  3254   &  2002-10-18  &  I    &  8.3    \\
RX~J0439.0+0520    &  527    &  2000-08-29  &  I    &  8.8    &    ~                  &  6109   &  2004-12-11  &  I    &  32.7   \\
~                  &  9369   &  2007-11-12  &  I    &  18.8   &    ~                  &  6172   &  2004-12-25  &  I    &  26.2   \\
~                  &  9761   &  2007-11-15  &  I    &  7.9    &    ~                  &  7720   &  2007-11-08  &  I    &  5.8    \\
Zwicky~2701        &  3195   &  2001-11-04  &  S    &  14.6   &    ~                  &  9379   &  2008-10-17  &  I    &  28.9   \\
~                  &  7706   &  2007-06-25  &  I    &  4.6    &    ~                  &  10785  &  2008-10-18  &  I    &  27.4   \\
~                  &  12903  &  2011-02-03  &  S    &  92.1   &    MACS~J1427.2+4407  &  6112   &  2005-02-12  &  I    &  8.4    \\
RX~J1504.1$-$0248  &  4935   &  2004-01-07  &  I    &  9.2    &    ~                  &  9380   &  2008-01-14  &  I    &  23.0   \\
~                  &  5793   &  2005-03-20  &  I    &  30.7   &    ~                  &  9808   &  2008-01-15  &  I    &  13.4   \\

%% file: tables/observation_info2.tex
MACS~J1427.2+4407  &  11694  &  2010-10-09  &  S    &  6.0    & ~                  &  14437  &  2012-09-16  &  I    &  23.1   \\
MACS~J1423.8+2404  &  1657   &  2001-06-01  &  I    &  16.2   & ~                  &  15572  &  2012-10-29  &  I    &  14.1   \\
~                  &  4195   &  2003-08-18  &  S    &  107.5  & ~                  &  15574  &  2012-10-31  &  I    &  11.3   \\
SPT~J2331$-$5051   &  9333   &  2009-08-12  &  I    &  26.4   & ~                  &  15579  &  2012-11-11  &  I    &  17.3   \\
~                  &  11738  &  2009-08-30  &  I    &  5.4    & ~                  &  15582  &  2012-11-17  &  I    &  17.3   \\
~                  &  18241  &  2016-08-29  &  I    &  79.1   & ~                  &  15588  &  2012-11-22  &  I    &  20.7   \\
~                  &  19697  &  2016-09-02  &  I    &  27.4   & ~                  &  15589  &  2012-11-24  &  I    &  9.9    \\
SPT~J2344$-$4242   &  13401  &  2011-09-19  &  I    &  10.7   & CL~J1415.2+3612    &  4163   &  2003-09-16  &  I    &  74.4   \\
~                  &  16135  &  2014-08-18  &  I    &  48.6   & ~                  &  12255  &  2010-08-30  &  S    &  60.4   \\
~                  &  16545  &  2014-08-20  &  I    &  51.6   & ~                  &  12256  &  2010-08-28  &  S    &  115.4  \\
SPT~J0000$-$5748   &  9335   &  2009-03-16  &  I    &  28.4   & ~                  &  13118  &  2010-09-01  &  S    &  44.6   \\
~                  &  18238  &  2016-08-26  &  I    &  118.2  & ~                  &  13119  &  2010-09-05  &  S    &  54.3   \\
~                  &  18239  &  2016-08-05  &  I    &  16.8   & 3C~186             &  3098   &  2002-05-16  &  S    &  16.9   \\
~                  &  19695  &  2016-08-24  &  I    &  23.7   & ~                  &  9407   &  2007-12-03  &  S    &  66.3   \\
SPT~J2043$-$5035   &  13478  &  2011-08-10  &  I    &  73.3   & ~                  &  9408   &  2007-12-11  &  S    &  39.6   \\
~                  &  18240  &  2016-08-05  &  I    &  96.9   & ~                  &  9774   &  2007-12-06  &  S    &  75.1   \\
SPT~J0615$-$5746   &  14017  &  2012-11-03  &  I    &  13.8   & ~                  &  9775   &  2007-12-08  &  S    &  15.9   \\
~                  &  14018  &  2012-09-15  &  I    &  31.2   & SPT~J2215$-$3537   &  22653  &  2020-08-17  &  I    &  32.3   \\
~                  &  14349  &  2012-11-09  &  I    &  21.8   & ~                  &  24614  &  2020-08-18  &  I    &  26.9   \\
~                  &  14350  &  2012-11-21  &  I    &  9.2    & ~                  &  24615  &  2020-08-20  &  I    &  8.2    \\
~                  &  14351  &  2012-11-12  &  I    &  22.5   \\

%% file: tables/fgas_parameters.tex
$\Upsilon_0$ & Gas depletion at $z=0$ & $\dunif(0.71, \, 0.87)^\star$ \\
$\Upsilon_1$ & Gas depletion evolution & $\dunif(-0.05, \, 0.05)^\star$ \\
$\alpha$ & Power-law slope of $\fgas$ with mass & $\dunif(-1, 1)$ \\
$\sigma_f$ & Intrinsic scatter of $\ln \fgas{}$ & $\dunif(0.0, \, 0.5)$ \\
$\eta_f$ & Power-law slope of shell $\fgas$ profile & $\dnorm(0.390, \, 0.024)^\star$ \\
$K_0$ & X-ray mass calibration at $z=0$ & $\dunif(0.0, \, 2.0)$ \\
$K_1$ & X-ray mass calibration evolution & $\dunif(-0.05,0.05)^\star$ \\
$\sigma_K$ & Lensing/X-ray mass ratio intrinsic scatter & $\dunif(0.0, 1.0)$ \\
$\eta_m$ & Power-law slope of mass profile & $\dnorm(1.065, \, 0.016)^\star$ \\
$W_0$ & Lensing mass bias at $z=0.4$ & $\dnorm(0.96, \, 0.09)^\star$ \\
$W_1$ & Lensing mass bias evolution & $\dnorm(-0.09, \, 0.03)^\star$ \\

%% file: tables/astro_parameters.tex
$\Omegam$ & Total matter density normalized to $\rhocr$ & $\dunif(0.0, \, 1.0)$ \\
$\Omegab$ & Baryon density normalized to $\rhocr$ & $\dunif(0.0, \, 1.0)$ \\
$H_0$ & Hubble parameter in km\,s$^{-1}$\,Mpc$^{-1}$ & $\dunif(20, \, 200)$ / $\dnorm(72, \, 8)^\dagger$ \\
$\Omegak$ & Equivalent energy density due to curvature & $\dunif(-0.65, \, 0.65)$ \\
$w$ & Dark energy equation of state & $\dunif(-5, \, 5)$ \\
$\Omegab h^2$ & Baryon density (derived) & $\dnorm(0.0215, \, 0.0005)^\dagger$ \\

%% file: tables/theta_parameters.tex
$\Omegab h^2$ & Baryon density & $\dunif(0.005, \, 0.1)$ \\
$\mysub{\Omega}{c} h^2$ & Cold dark matter density & $\dunif(0.001, \, 0.99)$ \\
$\theta_\mathrm{s}$ & Angular size of the sound horizon at last scattering & $\dunif(0.5, \, 10)$ \\
$\Omega_k$ & Effective density from spatial curvature & $\dunif(-0.65, \, 0.65)$ \\
$w$ ($w_0$) & Dark energy equation of state (at $z=0$) & $\dunif(-5, \, 5)$ \\
$w_a$ & Evolution parameter for $w(a)$ & $\dunif(-10, \, 10)$ \\
$\tau$ & Optical depth to reionization & $\dunif(0.01, \, 0.8)$ \\
$\log\,10^{10}\mysub{A}{s}$ & Scalar power spectrum amplitude & $\dunif(1.61, \, 3.91)$ \\
$\mysub{n}{s}$ & Scalar spectral index & $\dunif(0.8, \, 1.2)$ \\
$\Sigma\,m_\nu$ & Species-summed (degenerate) neutrino mass in eV & $=0.056$ \\
$\mysub{N}{eff}$ & Effective number of neutrino species & $=3.046$ \\
$H_0$ & Hubble parameter in km\,s$^{-1}$\,Mpc$^{-1}$ (derived) & $\dunif(20, \, 200)$ \\